\documentclass[showpacs,showkeys, groupedaddress,twocolumn, floatfix,aps, prl,reprint]{revtex4-1}
\usepackage[english]{babel}
\usepackage{fontenc}
\usepackage{graphicx}
\usepackage{amsmath}
\usepackage{epstopdf}
\usepackage{amssymb}
\usepackage{amsbsy}
\usepackage{amscd}
\usepackage{xcolor}
\usepackage{bbm}
\usepackage{braket}
\usepackage{bm}
\usepackage{siunitx}
\usepackage[normalem]{ulem} 
\usepackage{verbatim}
\usepackage{url}
\usepackage{dsfont} 
\usepackage[]{physics}
\usepackage[verbose,hypertexnames=false,bookmarksopenlevel=1,filecolor=blue,
linkcolor=blue,citecolor=blue,urlcolor=blue,pdfstartview=FitH,bookmarksopen,bookmarksnumbered,
colorlinks,plainpages=false,linktocpage]{hyperref}
\DeclareMathAlphabet\mathbfcal{OMS}{cmsy}{b}{n}

\allowdisplaybreaks
\begin{document}

\newcommand{\ua}[1]{u_{\alpha}^{n_{#1}}}
\newcommand{\ub}[1]{u_{\beta}^{n_{#1}}}

\newcommand{\ea}{\varepsilon_{\alpha}}
\newcommand{\eb}{\varepsilon_{\beta}}
\newcommand{\ih}{\frac{1}{\hbar}}
\newcommand{\G}{\mathcal{G}}
\newcommand{\se}{\boldsymbol{\Sigma}}
\newcommand{\kx}{k_{x}}
\newcommand{\ky}{k_{y}}
\newcommand{\gellmann}[1]{\boldsymbol{\lambda_{#1}}}
\newcommand{\Vimp}{\mathcal{V}_{\mathrm{imp}}}

\newcommand{\f}[1]{\mathbf{#1}}
\newcommand{\bs}[1]{\boldsymbol{#1}}

\newcommand{\id}{\mathds{1}}
\newcommand{\vol}{\mathfrak{V}}

\title{Floquet-Drude Conductivity}
\author{Martin Wackerl}
\email{martin.wackerl@ur.de}
\author{Paul Wenk}
\email{paul.wenk@ur.de}
\author{John Schliemann}
\email{john.schliemann@ur.de}
\affiliation{Institute for Theoretical
  Physics, University of Regensburg, 93040 Regensburg, Germany}

\date{\today }
\begin{abstract}
  This letter presents a generalization of the Drude conductivity for
  systems which are exposed to periodic driving. The probe bias is
  treated perturbatively by using the Kubo formula, whereas the
  external driving is included non-perturbatively using the Floquet
  theory. Using a new type of four-times Green's functions disorder is
  approached diagrammatically, yielding a fully analytical expression
  for the Floquet-Drude conductivity. Furthermore, the Floquet Fermi's
  golden rule is generalized to $tt'$-Floquet states, connecting the
  Floquet-Dyson series with scattering theory for Floquet states. Our
  formalism allows for a direct application to numerous systems
  e.g. graphene or spin-orbit systems.
\end{abstract}

\maketitle
\begin{figure}[htbp]
  \centering
  \includegraphics[width=0.8\linewidth]{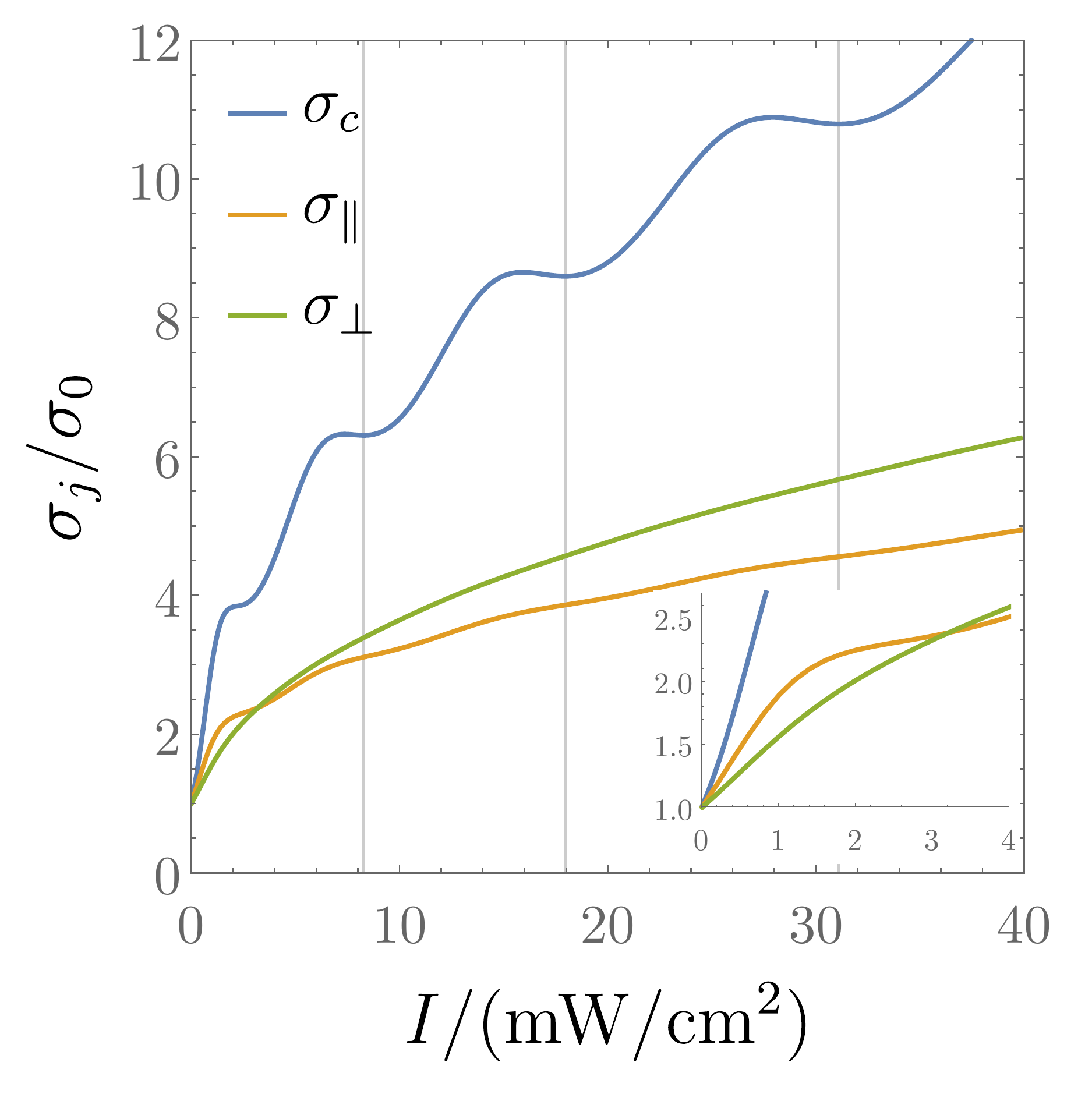}
  \caption{The Drude conductivity of a 2DEG irradiated by a circularly ($\sigma_c$) and
    linearly polarized light of intensity $I$ is plotted in comparison
    to the undriven case ($\sigma_0$). For linearly polarized light,
    $\sigma_\perp$  and $\sigma_\parallel$ are shown.
    The vertical lines
    indicate the minima of $\sigma_c$. The inset shows a
    magnification for small intensities. The
  ratio $\sqrt{E_F/\text{meV}}/(\omega/(\text{rad/s}))^2 = \sqrt{10}/10^{12}$
is chosen.}
  \label{fig:cond}
\end{figure}

%
%
\paragraph{Introduction.---\!\!\!\!\!}Paul Drude published his theory of electric
transport in metals as long ago as 1900 \cite{Drude00,Drude00_2},
which is today known as Drude theory. To the present day several
approaches have been developed to deepen the understanding of the
microscopic mechanisms occurring in charge transport, including
scattering theory using Fermi's golden rule~\cite{Dirac27,Fermi50} or
quantum corrections to the Drude conductivity. The latter covers weak
(anti-)localization~\cite{Altshuler1981a, Hikami80} in the form of
geometry or spin dependent
corrections~\cite{Chakravarty1986,Beenakker1988a, Iordanskii1994,
  Berman14, Knap1996, Meyer2002a, Wenk10, Kammermeier16,
  Kettemann07}. In contrast to studies of static systems the
development of lasers and masers generated a rising activity on
explicitly time-dependent Hamiltonians, where the external field
cannot be considered a small perturbation~\cite{Grifoni98}. In
the most recent decade, owing to the possibility of changing the topology
of a system by means of external driving, the investigation of
transport in driven systems increased~\cite{Kitagawa11,
  Zhou11, Oka, Morina15, Skvortsov1998, Shelykh15, Shi03, Dehghani15,
  Kibis14, Ganichev17}. This includes transport in driven
systems~\cite{Genske15,Esin2018}, either with or without
disorder~\cite{Paraj15,Paraj17,ParajPhd}, or the photo-voltaic Hall
effect \cite{Oka}. Most works studying the renormalization of
conductivity, due to an external driving, use a perturbative approach
regarding the external driving~\cite{Shelykh15,Morina15}. In this
letter we present a new general formalism that allows the
determination of the
Drude conductivity in the presence of a non-perturbative external
driving. We unify linear response theory and Floquet formalism to
account for the probe bias and an external driving. Using a new type
of four-times Green's function formalism we derive both a Floquet-Dyson
series and a generalized Floquet Fermi's golden rule. To prove the
consistency, both are shown to yield the same scattering time, a
link that was missing so far. Finally, we present a closed analytical
form for the Floquet-Drude conductivity and apply the developed theory
to a 2DEG with circularly polarized external driving. The analysis
shows that previous results have been overestimating the effect of the
driving on the conductivity.
%
%
\paragraph{Kubo formula for periodically driven systems.---\!\!\!\!\!}
Our first aim is to express the linear
conductivity using Floquet states~\cite{Floquet83,Grifoni98,Holthaus15} $|\psi_{\alpha}(t)\rangle  = \exp(-\frac{i}{\hbar}\ea t) | u_{\alpha}(t) \rangle$ ,
where $\alpha$ is labeling a discrete set of quantum numbers and $\ea$
are the corresponding quasienergies \cite{zel1967quasienergy,ritus1967shift}. The
periodicity of the Floquet functions, which are eigenstates of the Floquet Hamiltonian
$H_{F}(t) = H(t) -i\hbar\partial_{t}$, allow for the Fourier expansion
$| u_{\alpha}(t) \rangle = \sum_{n}e^{-in\Omega t} |u_{\alpha}^{n}
\rangle$ with $\Omega$ being the frequency of the external driving.
The probe bias, with the corresponding vector potential $\mathcal{A}(\f{q},\omega)$, is treated perturbatively in linear response theory,
i.e., using the Kubo formula in momentum space~\cite{Mahan2013}
\begin{align}
\begin{aligned}\label{Kubo}
   &\langle J^{a}(\f{q},\omega) \rangle ={}
   {} \frac{i}{\hbar \vol}\sum_b \int_{-\infty}^{\infty} dt'
   \int_{-\infty}^{\infty} dt\Big[ e^{i\omega t} \Theta(t-t') \\
   &\times\bigl\langle [J^{a}(\f{q},t),J^{b}(-\f{q},t')] \bigr\rangle \mathcal{A}^{b}(\f{q},t') \Big] -\frac{e^2 n}{m}\mathcal{A}^{a}(\f{q},\omega)\;,
\end{aligned}
\end{align}
with $a,b \in \{x,y,z\}$, $\omega$ being the frequency of the response
current, $\vol$ the volume of the system, $e<0$ the electron charge,
$\Theta(\cdot)$ the Heaviside function, $m$ the effective electron
mass, $n$ the electron density. $\langle\cdot\rangle$ denotes the
statistical average with respect to the system\rq s state which will
in the presence of external driving not be in equilibrium. However, in
what follows we shall assume the system to be in a stationary state so
that occupation numbers of Floquet states are
time-independent~\cite{Gupta03, Hsu06, Zhou11, Scholz13, Shirai15,
  Shirai16, Genske15}. Treating the external driving 
non-perturbatively, the current operators are expanded using Floquet
states $\ket{\psi_{\alpha}} = c_{\alpha}^{\dagger}\ket{0}$,
$J^{a,b}(\f{q},t)=\sum_{\alpha\beta}J^{a,b}_{\alpha\beta}(\f{q},t)c_{\alpha}^{\dagger}c_{\beta}$,
with the fermionic annihilation (creation) operators
$c_{\alpha}^{(\dagger)}$. The expansion coefficients are the matrix
elements
$J_{\alpha\beta}^{a,b}(\f{q},t)=\bra{\psi_{\alpha}(t)} J^{a,b}(\f{q})
\ket{\psi_{\beta}(t)}$. The current expectation value is no longer a
simple product of resistance and electric field
$\vb{E} = -\partial_t \mathbfcal{A}$ since it is convoluted over the
bias frequency $\omega'$,
\begin{align}\label{J_integral_sigma_E}
   \langle J^{a}(\f{q},\omega)\rangle = \sum_{b}
   \int_{-\infty}^{\infty}d\omega'\ 
   \bar{\sigma}^{ab}(\f{q},\omega,\omega')E^{b}(\f{q},\omega')\ .
\end{align}
Thus far, the conductivity tensor $\bar{\sigma}^{ab}$ depends on the
bias frequency, the momentum $\f{q}$ and on the resulting current
frequency $\omega$,
\begin{align}
\begin{aligned}\label{Sigma_bar}
  \bar{\sigma}^{ab}(\f{q},\omega,\omega') = & \frac{i}{\hbar\omega
    \vol}
  \sum_{\alpha\beta} \sum_{n_{1}..n_{4}=-\infty}^\infty (f_{\alpha}-f_{\beta})\\
  \times\ & \frac{\langle \ua{1} | j^{a}(\f{q}) | \ub{2} \rangle\langle \ub{3}  | j^{b}(-\f{q}) | \ua{4} \rangle }{\omega+\ih(\ea-\eb) + (n_{1}-n_{2})\Omega+i0^{+}}\\
  \times\ & \delta\bigl( \omega + (n_{1}-n_{2}+n_{3}-n_{4})\Omega - \omega' \bigr) \\
  +\ & i\frac{e^2 n }{m\omega} \delta_{ab}\delta(\omega-\omega')\ ,
\end{aligned}
\end{align}
with the single particle current operator $j^i$ and the distribution
function
$f_{\alpha,\beta} = \langle c_{\alpha,\beta}^{\dagger}c_{\alpha,\beta}
\rangle$. In what follows we limit our calculations to a position independent driving
field, i.e., $\f{q}={\bf 0}$. Since we are ultimately interested in
the DC conductivity, Eq.~\eqref{J_integral_sigma_E} simplifies to
$\langle J^{a}(\f{q},\omega)\rangle = \sum_{b}
\sigma^{ab}(\f{q},\omega)E^{b}(\f{q},\omega)$, as shown in the
supplemental material (SM). This allows us to express the real part of
the longitudinal DC conductivity as
\begin{align} \label{sigma_xx}
   \underset{\omega\to 0}{\mathrm{Re}}\,[\sigma^{xx}(0,\omega)] =
   \frac{\pi\hbar}{\vol} \left(\frac{e}{m}\right)^{2}
  \int\limits_{\lambda - \frac{\hbar\Omega}{2}}^{\lambda+\frac{\hbar\Omega}{2}}d\varepsilon
   \left(-\frac{\partial f}{\partial \varepsilon} \right) \sigma(\varepsilon) 
\end{align}
along with the definition
\begin{align}
\begin{aligned}
   \sigma(\varepsilon) = & \sum_{nn'=-\infty}^\infty\sum_{\alpha\beta} \bra{ u_{\alpha}^{n} } p^{x} \ket{ u_{\beta}^{n} } 
   \big\langle u_{\beta}^{n'} \big| p^{x} \big|  u_{\alpha}^{n'} \big\rangle\\ 
   & \times  \delta ( \varepsilon -\ea)\delta ( \varepsilon -\eb )\;,
\end{aligned}
\end{align}
and the abbreviation $\mathrm{Re}_{\omega\to 0}[\cdot]$ for
$\lim_{\omega\to 0}\mathrm{Re}[\cdot]$.  A similar result has already
been derived in Ref.~\cite{Zhou11} using the Keldysh framework. In
derivating Eq.~\eqref{sigma_xx} one requires the difference of two
quasienergies to always be smaller than the photon energy of the
external driving.
Thus far, we have not used the convenient choice of the
quasienergy~\cite{Bilitewski15} to be in
$[-\hbar\Omega/2,\hbar\Omega/2)$, however we must choose a
suitable, possibly momentum dependent function $\lambda$ such that
\begin{align}
  \forall_\alpha: \lambda - \frac{\hbar\Omega}{2} \le \varepsilon_{\alpha} < \lambda + \frac{\hbar\Omega}{2}\;.
\end{align}
%
%
%
\paragraph{Four-times Floquet Green's function and conductivity.---\!\!\!\!\!} In
this section we set up a formalism using four-times Green's functions
to express the result of the foregoing section for the conductivity in
terms of Green's functions. These are the building blocks for the
Floquet-Dyson equation. First, we define a $tt'$-state for the
$\ell$-th Floquet zone \cite{Martinez03,Grifoni98,Drese99,Sambe73}
\begin{align}\label{tt'state}
   \ket{\psi^{\ell}_{\alpha}(t,t')} = e^{-\frac{i}{\hbar}(\ea+\ell\hbar\Omega)t} \ket{u_{\alpha}(t')}e^{i\ell\Omega t'}
\end{align}
recovering for $t=t'$ the Floquet state solution
$|\psi_{\alpha}(t)\rangle$ of the time-dependent Schr\"odinger
equation. From these a bare four-times Green's function is constructed
\begin{align}
\begin{aligned} \label{tt'Green}
   \G^{r,a}_{0}(t_{1},t_{2},t_{1}',t_{2}') = \mp i\Theta(\pm(t_{1}-t_{2})) \\ \times\frac{1}{T}\sum_{\ell=-\infty}^{\infty}\sum_{\alpha}\ket{\psi^{\ell}_{\alpha}(t_{1},t_{1}')}\bra{\psi^{\ell}_{\alpha}(t_{2},t_{2}')}
\end{aligned}
\end{align}
with $T=2\pi/\Omega$. This propagator fulfils
\begin{align}
\begin{aligned} \label{GreensDef}
   & \Big( i\partial_{t_{1}}-\tfrac{1}{\hbar} H_{F}(t_{1}') \Big)
   \G^{r,a}_{0}(t_{1},t_{2},t_{1}',t_{2}') =\\
   &\delta(t_{1}-t_{2}) \sum_{s=-\infty}^{\infty}\delta(t_{1}'-t_{2}'+sT)  \id\;.
\end{aligned}
\end{align}
Fourier transforming the Green's function and expanding the Floquet
function into a Fourier series yields
\begin{align}
\begin{aligned}\label{Green_series}
   \G^{r,a}_{0}(\varepsilon,t_{1}',t_{2}') = \frac{1}{T} \sum_{n,n'=-\infty}^{\infty} \G_{0}^{r,a}(\varepsilon,n,n') \\ \times e^{-in\Omega t_{1}'}e^{in'\Omega t_{2}'}\ .
\end{aligned}
\end{align}
Generalizing the completeness relation of the Floquet functions to different times
\begin{align}
	\begin{aligned}
		\sum_{\alpha} |u_{\alpha}(t_{1}')\rangle \langle u_{\alpha}(t_{2}')| \sum_{\ell=-\infty}^{\infty}\delta(t_{1}'-t_{2}'+\ell T)\\
		= \id\sum_{\ell=-\infty}^{\infty}\delta(t_{1}'-t_{2}'+\ell T)
	\end{aligned}
\end{align}
gives particular insight to the Lehmann representation of the four-times Floquet Green's function:
\begin{align}\label{FloquetLehmann}
	\G_{0}^{r,a}(\varepsilon,t_{1}',t_{2}') = \frac{\id\sum_{\ell=-\infty}^{\infty}\delta(t_{1}'-t_{2}'+\ell T)}{\ih\varepsilon-\ih H_{F}(t_{1}')}\ .
\end{align}
One can show that the Fourier coefficients \cite{Martinez05,Martinez06}
\begin{align} 
   \G^{r,a}_{0}(\varepsilon,n,n') = \sum_{\ell=-\infty}^{\infty}\sum_{\alpha} \frac{ \big| u_{\alpha}^{n+\ell}\big\rangle \big\langle u_{\alpha}^{n'+\ell}\big|}{\frac{1}{\hbar}\varepsilon-\frac{1}{\hbar}\ea - \ell\Omega \pm i0^{+}}\,,
\end{align}
%
%
%
are equal to the inverse of the Floquet matrix
\cite{Eissing16,Eissing16B,Taberner17,Aoki14,Paraj15,Paraj17,Rentrop14,Genske15,Wu08}. The
periodicity of the Floquet eigenstates \cite{Shirley63,Shirley65}
suggests defining the unitary transformation $\bs{T}$ as in Sec. V of
Ref.~\cite{Tsuji08} which diagonalizes the Green's function
\begin{align} \label{eigengreen}
   \big(\bs{D}(\varepsilon)\big)^{nn'}_{\alpha\beta} &\equiv \left( \bs{T}^{\dagger}\bs{G}_{0}^{r,a}(\varepsilon) \bs{T} \right)^{nn'}_{\alpha\beta} \\
   & = \frac{\delta_{\alpha\beta}\delta_{nn'}}{\ih\varepsilon-\ih\ea - n \Omega\pm i0^{+}}
\end{align}
where we denoted the matrix spanned by the Fourier components of the Green's function as
$\big(\bs{G}_{0}^{r,a}(\varepsilon)\big)_{nn'} = \G^{r,a}_{0}(\varepsilon,n,n')$ .
The Green's function defined in Eq.~\eqref{Green_series} is used to express the conductivity from Eq.~\eqref{sigma_xx} as
\begin{align} 
\begin{aligned} \label{Sigma_bare}
   \underset{\omega\to 0}{\mathrm{Re}} [\sigma^{xx}(0,\omega)] & =  \frac{-1}{4\pi\hbar \vol} \Big(\frac{e}{m}\Big)^{2} \\
   \times & \int\limits^{\lambda+\frac{\hbar\Omega}{2}}_{\lambda-\frac{\hbar\Omega}{2}}d\varepsilon \left(-\frac{\partial f}{\partial \varepsilon} \right) \int\limits_{0}^{T} dt_{1}' \int\limits_{0}^{T} dt_{2}' \\
   \times & \mathrm{tr} \biggl[ p^{x} \Bigl( \mathcal{G}^{r}_{0}(\varepsilon,t_{1}',t_{2}') - \mathcal{G}^{a}_{0}(\varepsilon,t_{1}',t_{2}') \Bigr) \\
   & \hspace{0.45cm} p^{x} \Bigl( \mathcal{G}^{r}_{0}(\varepsilon,t_{2}',t_{1}') - \mathcal{G}^{a}_{0}(\varepsilon,t_{2}',t_{1}') \Bigr) \biggr]\ .
\end{aligned}
\end{align}
%
%
%
\paragraph{Floquet-Dyson equation and generalized Floquet Fermi's
  golden rule.---\!\!\!\!\!} The focus of this section is to formulate a
perturbative approach to include disorder in the expression of the
conductivity described by bare propagators, i.e.,
Eq.~\eqref{Sigma_bare}.  In the following we will use the notation
\begin{align}\label{PosNotation}
\begin{aligned}
   \bra{\f{x}}\G^{r,a}(t_{1},t_{2},t_{1}',t_{2}')\ket{\f{x}'} & \equiv \G^{r,a}(t_{1},t_{2},\f{x},\f{x'},t_{1}',t_{2}')\;,\\
   \bra{\f{x}}\G^{r,a}(\varepsilon,t_{1}',t_{2}')\ket{\f{x}'} & \equiv \G^{r,a}(\varepsilon,\f{x},\f{x'},t_{1}',t_{2}')
 \end{aligned}
\end{align}
for the matrix elements of the Green's function in real space.  The
Green's function for the system with an impurity potential $V(\f{x}_{1},t_{1},t_{1}')$ at site $\f{x}_1$ is supposed to fulfill
\begin{align}
\begin{aligned} \label{Gp}
   \Big( i\partial_{t_{1}}-\tfrac{1}{\hbar} \mathcal{H}_{F}(\f{x}_{1},t_{1},t_{1}')\Big) \G^{r,a}_{\mathrm{p}}(t_{1},t_{2},\f{x}_{1},\f{x}_{2},t_{1}',t_{2}') = \\ \delta(t_{1}-t_{2})\delta(\f{x}_{1}-\f{x}_{2})\sum_{s=-\infty}^{\infty}\delta(t_{1}'-t_{2}'+sT)  \id
\end{aligned}
\end{align} 
with
$\mathcal{H}_{F}(\f{x}_{1},t_{1},t_{1}') =
H_{F}(\f{x}_{1},t_{1}')+V(\f{x}_{1},t_{1},t_{1}')$. We limit the
calculation presented here to the time-independent potential
$V(\f{x})$, as including an explicit time-dependence is
straightforward (see SM). Following the standard
steps~\cite{BruusFlensberg,Mahan2013,akkermans,rammer}, we can derive a
recursive integral expansion, i.e., a Dyson series, for the Green's
function, in case of a system perturbed by impurities
\begin{align}
\begin{aligned} \label{Dyson-series}
   \G^{r,a}_{\mathrm{p}}(\varepsilon,\f{x}_{1},\f{x}_{2},n,n') =\ &  \G^{r,a}_{0}(\varepsilon,\f{x}_{1},\f{x}_{2},n,n')\\
   +\,\ih\int_{V_{\f{x}}} d\f{x} \sum_{n_{1}=-\infty}^{\infty}\ & \G^{r,a}_{\mathrm{p}}(\varepsilon,\f{x}_{1},\f{x},n,n_{1}) V(\f{x}) \\
   \times & \G_{0}^{r,a}(\varepsilon,\f{x},\f{x}_{2},n_{1},n')\ .
\end{aligned}
\end{align}
The impurity potential $V(\f{x})= \sum_{i}^{N_{\mathrm{imp}}}
v(\f{x}-\f{r}_{i})$ is assumed to be a Gaussian random potential, which
is uncorrelated such that the impurity
average yields $\langle v(\f{x})v(\f{x}') \rangle_{\mathrm{imp}} =
\Vimp\delta(\f{x}-\f{x}')$ $\Leftrightarrow$ $\nu(\f{k}) = \Vimp$ and $\langle v(\f{x})
\rangle_{\mathrm{imp}}  = 0$ (white noise)~\cite{akkermans,rammer}.
%
%
%
Fourier transforming Eq.~\eqref{Dyson-series} into momentum space and
performing a disorder average one arrives at the expression for the
disorder averaged Green's function
\begin{align}
\begin{aligned}\label{RecursiveGreen}
   \bs{G}^{r,a}& (\varepsilon,\f{k}) = \ \bs{G}_{0}^{r,a}(\varepsilon,\f{k}) \\
     & + \bs{G}^{r,a}_{0}(\varepsilon,\f{k}) \sum_{n=1}^{\infty} \bigg[ \se^{r,a}(\varepsilon,\f{k})\bs{G}^{r,a}_{0}(\varepsilon,\f{k})\biggr]^{n}
\end{aligned}
\end{align}
where the self-energy $\se^{r,a}(\varepsilon,\f{k})$ is the sum over
all irreducible diagrams.
%
Applying the transformation $\bs{T}$, the solution of the recursive
Eq.~\eqref{RecursiveGreen} in the eigenbasis is governed by
\begin{align}
\begin{aligned}
   \bs{T}^{\dagger}(\f{k})\bs{G}^{r,a}(\varepsilon,\f{k})\bs{T}(\f{k})& = \bigg[ \bs{D}(\varepsilon,\f{k}) \\ 
   &+ \bs{T}^{\dagger}(\f{k})\se^{r,a}(\varepsilon,\f{k})\bs{T}(\f{k})  \bigg]^{-1}
\end{aligned}
\end{align}
with the diagonal matrix $\bs{D}(\varepsilon,\f{k})$ given in Eq.~\eqref{eigengreen}.
%
%
%
%
The difference of the retarded and advanced self-energy in the first order
Born approximation (1BA) can be related to a scattering time derived within
the framework of the Floquet Fermi's Golden rule for
$tt'$-states~\eqref{tt'state} as
\begin{align} \label{ScatteringTimett'states}
\begin{aligned}
  \left(\frac{1}{\bs{\tau}(\varepsilon,\f{k})}
  \right)_{\alpha\beta}^{nn'} ={}&  i \Big( \bs{T}^{\dagger}(\f{k}) \big(
  \se^{r}_{1\mathrm{BA}}(\varepsilon,\f{k}) \\
 {}& -\,\se^{a}_{1\mathrm{BA}}(\varepsilon,\f{k})\big)\bs{T}(\f{k})
  \Big)_{\alpha\beta}^{nn'}\;.
\end{aligned}
\end{align}
To our knowledge, this remarkable connection has not been established
before. The related scattering rates are given by
\begin{align}
  \Gamma_{\alpha\beta}^{nn'}(\f{k},\f{k}') 
  ={}&  \frac{2\pi}{\hbar}V_{\f{kk}'}^{2} \sum_{\gamma}
    c_{\alpha\gamma}^{n}(\f{k},\f{k'})
    \big(c_{\beta\gamma}^{n'}(\f{k},\f{k'})
    \big)^{*}\nonumber\\
  {}& \times\delta \big( \varepsilon-\varepsilon_{\gamma}(\f{k}') \big)\label{scatteringNN}
\end{align}
with $ c_{\alpha\beta}^{n}(\f{k},\f{k'}) \equiv \sum_{m=-\infty}^{\infty}  \big(u_{\alpha}^{m+n}(\f{k})\big)^{*} u_{\beta}^{m}(\f{k}')$.
One can provide a connection to previous studies by setting $n=n'=0$
in Eq.~\eqref{scatteringNN}, which results in the scattering rates
given in Ref.~\cite{Kitagawa11,Bilitewski15}. However, in general the
relaxation rate matrix is not diagonal in the Floquet space nor can it
be reduced to the $(nn')=(00)$ element only, as will be discussed later
on. Also, one should notice that only on the diagonal is the
difference of the retarded and advanced self energy equal to the
imaginary part of the retarded self-energy.
%
%
\paragraph{Floquet-Drude conductivity.---\!\!\!\!\!} Now, let us consider only the
self-energy corrections during the disorder average and perform the
time integration in Eq.~\eqref{Sigma_bare}. The disorder is not supposed
to change the eigenenergies of the bare system, hence we drop all
off-diagonal elements of the self energy. This allows us to proceed
analytically and ultimately express the conductivity in a compact way,
\begin{align}
\begin{aligned}
   &\underset{\omega\to 0}{\text{Re}}\, [\sigma^{xx}(0,\omega)] = \frac{\hbar}{4\pi \vol}\left(\frac{e}{m}\right)^{2} \int\limits_{-\hbar\Omega/2}^{\hbar\Omega/2}d\varepsilon \left(-\frac{\partial f}{\partial \varepsilon}\right) \\
   &\times\frac{1}{\vol_{\f{k}}}\sum_{\f{k}} k_{x}^{2}\sum_{\alpha}\sum_{n=-\infty}^{\infty} \Big(\Big( \frac{1}{2\bs{\tau}(\varepsilon,\f{k})} \Big)_{\alpha\alpha}^{nn}\Big)^{2}\\
   &\times\Big[ \Big( \tfrac{1}{\hbar}\varepsilon - \tfrac{1}{\hbar}\ea - n\Omega \Big)^{2} + \Big(\Big( \frac{1}{2\bs{\tau}(\varepsilon,\f{k})} \Big)_{\alpha\alpha}^{nn}\Big)^{2} \Big]^{-2}\label{condxx}\;.
\end{aligned}
\end{align}
%
%
%
%
\paragraph{Application to a 2DEG.---\!\!\!\!\!} For a simple non-trivial
application of the method described above we chose a 2DEG from a
direct semiconductor close to the $\Gamma$-point under illumination
with circularly polarized light. The effective model for the lowest
s-type conduction band is given by a parabolic Hamiltonian
$H = \f{p}^2/2m$.
The vector potential of the radiation
\begin{align}
   \f{A}(t) = A \cos(\Omega t)\f{\hat{x}}+A \sin(\Omega t)\f{\hat y}
\end{align}
is coupled to the momentum via minimal coupling leading to the time-dependent Hamiltonian 
\begin{align}
\begin{aligned}
   H(t) = \frac{\hbar^2}{2m}\Big[ \f{k}^2 + \gamma^{2} + 2 \gamma
   \big(k_{x}\cos(\Omega t) +  k_{y}\sin(\Omega t)\big) \Big] 
\end{aligned}
\end{align}
with the light parameter $\gamma \equiv eA/\hbar$. The solution of the time-dependent Schr\"odinger equation
\begin{align}
   i\hbar\frac{\partial}{\partial t} \Psi(\f{k},t) = H(t)\Psi(\f{k},t)
\end{align}
was already given by Kibis \cite{Kibis14}
\begin{align}\label{Psiexact}
   \Psi(\f{k},t) = e^{-\frac{i}{\hbar} \epsilon_{\f{k}}t } \sum_{n=-\infty}^{\infty} u_{n}(\f{k}) e^{-in\Omega t}
\end{align}
where the quasienergy and the Fourier components are given by
\begin{align}
   \epsilon_{\f{k}} &= \frac{\hbar^2}{2m}\left(\f{k}^2+\gamma^2\right) , \\ u_{n}(\f{k}) &= J_{n}\left(\frac{\hbar^2\gamma k}{2m}\Big/\hbar\Omega\right)e^{-in(\phi+\pi/2)}
\end{align}
with $\f{k} = k(\cos(\phi),\sin(\phi))$. As one can demonstrate by
solving the driven tight binding model for the square lattice with
lattice constant $a$ (see SM), the effective mass $m$ is also $\gamma$
dependent. Thus, we change $m\rightarrow m(\gamma)= m/J_0(a\gamma)$.
To evaluate the expression for the conductivity, one has to specify
the distribution function further. In the off resonant regime,
absorption of photons is suppressed, hence, a Fermi distribution can
be assumed. However, since the parabolic spectrum is unbounded, it is
not obvious how to set the Fermi energy $\varepsilon_{F}$ for the
driven parabolic spectrum \cite{Bilitewski15}. Here, we truncate the
momentum range to set the Fermi energy (for further discussion see
SM). Evaluating Eq.~\eqref{ScatteringTimett'states} yields for the
scattering time on the diagonal
\begin{align}
  \left(\frac{1}{\bs{\tau}(\varepsilon_F,\f{k})}\right)^{nn} = \frac{\Vimp m(\gamma)}{\hbar^{3}} \sum_{m=-\infty}^{\infty}J_{m+n}^{2}(z_{k})J_{m}^{2}(z_{\varepsilon})\;,\label{tauNN}
\end{align}
together with
\begin{align}
   z_{k} = \frac{\hbar^{2}\gamma k}{m(\gamma)}\Big/\hbar\Omega\ , \quad
  z_{\varepsilon} = \frac{\hbar^{2}\gamma \sqrt{2m(\gamma)\varepsilon_{F}/\hbar^{2}}}{m(\gamma)}\Big/\hbar\Omega\;.
\end{align}
We can further disregard all pairings of only retarded or only advanced
Green's functions in Eq.~\eqref{condxx}, since they give a contribution
of the order of $1/(\varepsilon_F \tau_0)$ with $\tau_{0}$ being the
scattering time of the undriven system \cite{rammer}. Aside from the
relation between the Fermi energy and the relaxation rate, in
the driven case one finds an important relation between relaxation rate and
driving frequency. This ratio controls whether or not only the $n=0$
element in Eq.~\eqref{tauNN} is significant: If $\Omega\tau_{0}\gg 1$,
the broadening of the nonzero Floquet modes is small enough such that
the leaking into the central Floquet zone is negligibly small. The
theory presented here also makes the regime accessible where
$\Omega\tau_{0}\simeq 1$. In that case, the nonzero Floquet modes
contribute significantly (for further details see SM). But even in
the off resonant regime, $\Omega\tau_{0}\gg 1$, previous
studies~\cite{Morina15} overestimate the effect of circular
driving as will be shown in the following.  The central entry of the
product of the retarded Green's function with an advanced one is
\begin{align}
\begin{split}
   & \big[\bs{G}^{r}(\varepsilon_{F},\f{k})\bs{G}^{a}(\varepsilon_{F},\f{k})\big]^{00} \\
   &= \frac{1}{\big( \ih\varepsilon_{F} - \ih\epsilon_{\f{k}} \big)^{2} + \big(\big(\frac{1}{2\bs{\tau}(\varepsilon_{F},\f{k})}\big)^{00}\big)^{2} }
\end{split} \\
   &\approx 2\pi\hbar\big(\bs{\tau}(\varepsilon_{F},\f{k})\big)^{00}\delta(\varepsilon_{F}-\epsilon_{\f{k}})\;.
\end{align}
Applying these simplifications to Eq.~\eqref{condxx}, one arrives at the conductivity
\begin{align}
\begin{aligned}\label{Sigma_gr_ga}
  \underset{\omega\to 0}{\text{Re}}\,[\sigma^{xx}(0,\omega)] =\,
  \frac{1}{\vol 4\pi}\left( \frac{e^2}{m(\gamma)}  \right) k_F^{2}\big(\bs{\tau}(\varepsilon_{F},k_F)\big)^{00}\;,
\end{aligned}
\end{align}
where the scattering time is evaluated at the Fermi energy and Fermi
wave vector $k_F = \sqrt{2m(\gamma)\varepsilon_{F}}/\hbar$,
\begin{align}
   \big(\bs{\tau}(\varepsilon_{F},k_{F})\big)^{00} = \left(\frac{\Vimp m(\gamma)}{\hbar^{3}}\sum_{l=-\infty}^{\infty}J_{l}^{4}(z_{\varepsilon}) \right)^{-1}\;.
\end{align}
Hence, the ratio between conductivity without driving and dressed
conductivity is given by
\begin{align}
  \frac{ \underset{\omega\to 0}{\text{Re}}\,[\sigma^{xx}(0,\omega)]}{
  \underset{\omega\to 0}{\text{Re}}\,
  [\sigma^{xx}(0,\omega)]\big|_{\gamma=0}} = \frac{J_0(a\gamma)}{\sum_{l=-\infty}^{\infty}J_{l}^{4}(z_{\varepsilon})}\;.
\end{align}
In Fig.~\ref{fig:cond} we present the conductivity of a 2DEG
irradiated by a circularly ($\sigma_c$) polarized light of intensity
$I$. For comparison also the results in case of linearly polarized
light are shown.  The applied approximations above allow for a direct
comparison with existing theories on the topic of conductivity in
driven systems. The central entry of the scattering time used in
Eq.~\eqref{Sigma_gr_ga} is equal to the result which one yields from
the Floquet Fermi's golden rule~\cite{Kitagawa11,Bilitewski15,Kibis14}
(proof in the SM). It is used e.g. by the authors of
Ref.~\cite{Morina15} to calculate the conductivity. However, the
equation \textit{ibid} overestimates the effect of the driving for the
latter.
%
\acknowledgments
This work was supported by the Deutsche
Forschungsgemeinschaft via GRK 1570 and project 336985961.
\bibliographystyle{apsrev4-1}
\bibliography{paper}

\def\url#1{}
\begin{thebibliography}{59}%
\makeatletter
\providecommand \@ifxundefined [1]{%
 \@ifx{#1\undefined}
}%
\providecommand \@ifnum [1]{%
 \ifnum #1\expandafter \@firstoftwo
 \else \expandafter \@secondoftwo
 \fi
}%
\providecommand \@ifx [1]{%
 \ifx #1\expandafter \@firstoftwo
 \else \expandafter \@secondoftwo
 \fi
}%
\providecommand \natexlab [1]{#1}%
\providecommand \enquote  [1]{``#1''}%
\providecommand \bibnamefont  [1]{#1}%
\providecommand \bibfnamefont [1]{#1}%
\providecommand \citenamefont [1]{#1}%
\providecommand \href@noop [0]{\@secondoftwo}%
\providecommand \href [0]{\begingroup \@sanitize@url \@href}%
\providecommand \@href[1]{\@@startlink{#1}\@@href}%
\providecommand \@@href[1]{\endgroup#1\@@endlink}%
\providecommand \@sanitize@url [0]{\catcode `\\12\catcode `\$12\catcode
  `\&12\catcode `\#12\catcode `\^12\catcode `\_12\catcode `\%12\relax}%
\providecommand \@@startlink[1]{}%
\providecommand \@@endlink[0]{}%
\providecommand \url  [0]{\begingroup\@sanitize@url \@url }%
\providecommand \@url [1]{\endgroup\@href {#1}{\urlprefix }}%
\providecommand \urlprefix  [0]{URL }%
\providecommand \Eprint [0]{\href }%
\providecommand \doibase [0]{http://dx.doi.org/}%
\providecommand \selectlanguage [0]{\@gobble}%
\providecommand \bibinfo  [0]{\@secondoftwo}%
\providecommand \bibfield  [0]{\@secondoftwo}%
\providecommand \translation [1]{[#1]}%
\providecommand \BibitemOpen [0]{}%
\providecommand \bibitemStop [0]{}%
\providecommand \bibitemNoStop [0]{.\EOS\space}%
\providecommand \EOS [0]{\spacefactor3000\relax}%
\providecommand \BibitemShut  [1]{\csname bibitem#1\endcsname}%
\let\auto@bib@innerbib\@empty
\bibitem [{\citenamefont {Drude}(1900{\natexlab{a}})}]{Drude00}%
  \BibitemOpen
  \bibfield  {author} {\bibinfo {author} {\bibfnamefont {P.}~\bibnamefont
  {Drude}},\ }\href {\doibase 10.1002/andp.19003060312} {\bibfield  {journal}
  {\bibinfo  {journal} {Annalen der Physik}\ }\textbf {\bibinfo {volume}
  {306}},\ \bibinfo {pages} {566} (\bibinfo {year}
  {1900}{\natexlab{a}})}\BibitemShut {NoStop}%
\bibitem [{\citenamefont {Drude}(1900{\natexlab{b}})}]{Drude00_2}%
  \BibitemOpen
  \bibfield  {author} {\bibinfo {author} {\bibfnamefont {P.}~\bibnamefont
  {Drude}},\ }\href
  {https://archive.org/stream/bub_gb_QFRMAAAAMAAJ#page/n177/mode/2up}
  {\bibfield  {journal} {\bibinfo  {journal} {Physikalische Zeitschrift}\
  }\textbf {\bibinfo {volume} {14}},\ \bibinfo {pages} {161} (\bibinfo {year}
  {1900}{\natexlab{b}})}\BibitemShut {NoStop}%
\bibitem [{\citenamefont {Dirac}(1927)}]{Dirac27}%
  \BibitemOpen
  \bibfield  {author} {\bibinfo {author} {\bibfnamefont {P.~A.~M.}\
  \bibnamefont {Dirac}},\ }\href {\doibase 10.1098/rspa.1927.0039} {\bibfield
  {journal} {\bibinfo  {journal} {Proceedings of the Royal Society A:
  Mathematical, Physical and Engineering Sciences}\ }\textbf {\bibinfo {volume}
  {114}},\ \bibinfo {pages} {243} (\bibinfo {year} {1927})}\BibitemShut
  {NoStop}%
\bibitem [{\citenamefont {Fermi}(1974)}]{Fermi50}%
  \BibitemOpen
  \bibfield  {author} {\bibinfo {author} {\bibfnamefont {E.}~\bibnamefont
  {Fermi}},\ }\href {https://books.google.de/books?id=WQtkYCWTcicC} {\emph
  {\bibinfo {title} {Nuclear Physics: A Course Given by Enrico Fermi at the
  University of Chicago}}},\ Midway reprint\ (\bibinfo  {publisher} {University
  of Chicago Press},\ \bibinfo {year} {1974})\BibitemShut {NoStop}%
\bibitem [{\citenamefont {Al'tshuler}\ \emph {et~al.}(1981)\citenamefont
  {Al'tshuler}, \citenamefont {Aronov}, \citenamefont {Larkin},\ and\
  \citenamefont {Khmel'nitskii}}]{Altshuler1981a}%
  \BibitemOpen
  \bibfield  {author} {\bibinfo {author} {\bibfnamefont {B.~L.}\ \bibnamefont
  {Al'tshuler}}, \bibinfo {author} {\bibfnamefont {A.~G.}\ \bibnamefont
  {Aronov}}, \bibinfo {author} {\bibfnamefont {A.~I.}\ \bibnamefont {Larkin}},
  \ and\ \bibinfo {author} {\bibfnamefont {D.~E.}\ \bibnamefont
  {Khmel'nitskii}},\ }\href
  {http://www.jetp.ac.ru/cgi-bin/e/index/e/54/2/p411?a=list} {\bibfield
  {journal} {\bibinfo  {journal} {Sov. Phys. JETP}\ }\textbf {\bibinfo {volume}
  {54}},\ \bibinfo {pages} {0411} (\bibinfo {year} {1981})}\BibitemShut
  {NoStop}%
\bibitem [{\citenamefont {Hikami}\ \emph {et~al.}(1980)\citenamefont {Hikami},
  \citenamefont {Larkin},\ and\ \citenamefont {Nagaoka}}]{Hikami80}%
  \BibitemOpen
  \bibfield  {author} {\bibinfo {author} {\bibfnamefont {S.}~\bibnamefont
  {Hikami}}, \bibinfo {author} {\bibfnamefont {A.~I.}\ \bibnamefont {Larkin}},
  \ and\ \bibinfo {author} {\bibfnamefont {Y.}~\bibnamefont {Nagaoka}},\ }\href
  {\doibase 10.1143/PTP.63.707} {\bibfield  {journal} {\bibinfo  {journal}
  {Progress of Theoretical Physics}\ }\textbf {\bibinfo {volume} {63}},\
  \bibinfo {pages} {707} (\bibinfo {year} {1980})}\BibitemShut {NoStop}%
\bibitem [{\citenamefont {Chakravarty}\ and\ \citenamefont
  {Schmid}(1986)}]{Chakravarty1986}%
  \BibitemOpen
  \bibfield  {author} {\bibinfo {author} {\bibfnamefont {S.}~\bibnamefont
  {Chakravarty}}\ and\ \bibinfo {author} {\bibfnamefont {A.}~\bibnamefont
  {Schmid}},\ }\href {\doibase 10.1016/0370-1573(86)90027-X} {\bibfield
  {journal} {\bibinfo  {journal} {Physics Reports}\ }\textbf {\bibinfo {volume}
  {140}},\ \bibinfo {pages} {193} (\bibinfo {year} {1986})}\BibitemShut
  {NoStop}%
\bibitem [{\citenamefont {Beenakker}\ and\ \citenamefont {van
  Houten}(1988)}]{Beenakker1988a}%
  \BibitemOpen
  \bibfield  {author} {\bibinfo {author} {\bibfnamefont {C.~W.~J.}\
  \bibnamefont {Beenakker}}\ and\ \bibinfo {author} {\bibfnamefont
  {H.}~\bibnamefont {van Houten}},\ }\href {\doibase 10.1103/PhysRevB.38.3232}
  {\bibfield  {journal} {\bibinfo  {journal} {Physical Review B}\ }\textbf
  {\bibinfo {volume} {38}},\ \bibinfo {pages} {3232} (\bibinfo {year}
  {1988})}\BibitemShut {NoStop}%
\bibitem [{\citenamefont {Iordanskii}\ \emph {et~al.}(1994)\citenamefont
  {Iordanskii}, \citenamefont {Lyandageller},\ and\ \citenamefont
  {Pikus}}]{Iordanskii1994}%
  \BibitemOpen
  \bibfield  {author} {\bibinfo {author} {\bibfnamefont {S.~V.}\ \bibnamefont
  {Iordanskii}}, \bibinfo {author} {\bibfnamefont {Y.~B.}\ \bibnamefont
  {Lyandageller}}, \ and\ \bibinfo {author} {\bibfnamefont {G.~E.}\
  \bibnamefont {Pikus}},\ }\href
  {http://www.jetpletters.ac.ru/ps/1323/article_20010.shtml} {\bibfield
  {journal} {\bibinfo  {journal} {Jetp Letters}\ }\textbf {\bibinfo {volume}
  {60}},\ \bibinfo {pages} {206} (\bibinfo {year} {1994})}\BibitemShut
  {NoStop}%
\bibitem [{\citenamefont {Berman}\ \emph {et~al.}(2014)\citenamefont {Berman},
  \citenamefont {Khodas},\ and\ \citenamefont {Flatt{\'{e}}}}]{Berman14}%
  \BibitemOpen
  \bibfield  {author} {\bibinfo {author} {\bibfnamefont {D.}~\bibnamefont
  {Berman}}, \bibinfo {author} {\bibfnamefont {M.}~\bibnamefont {Khodas}}, \
  and\ \bibinfo {author} {\bibfnamefont {M.}~\bibnamefont {Flatt{\'{e}}}},\
  }\href {\doibase 10.1103/PhysRevX.4.011048} {\bibfield  {journal} {\bibinfo
  {journal} {Phys. Rev. X}\ }\textbf {\bibinfo {volume} {4}},\ \bibinfo {pages}
  {011048} (\bibinfo {year} {2014})}\BibitemShut {NoStop}%
\bibitem [{\citenamefont {Knap}\ \emph {et~al.}(1996)\citenamefont {Knap},
  \citenamefont {Skierbiszewski}, \citenamefont {Zduniak}, \citenamefont
  {Litwin-Staszewska}, \citenamefont {Bertho}, \citenamefont {Kobbi},
  \citenamefont {Robert}, \citenamefont {Pikus}, \citenamefont {Pikus},
  \citenamefont {Iordanskii}, \citenamefont {Mosser}, \citenamefont
  {Zekentes},\ and\ \citenamefont {Lyanda-Geller}}]{Knap1996}%
  \BibitemOpen
  \bibfield  {author} {\bibinfo {author} {\bibfnamefont {W.}~\bibnamefont
  {Knap}}, \bibinfo {author} {\bibfnamefont {C.}~\bibnamefont
  {Skierbiszewski}}, \bibinfo {author} {\bibfnamefont {A.}~\bibnamefont
  {Zduniak}}, \bibinfo {author} {\bibfnamefont {E.}~\bibnamefont
  {Litwin-Staszewska}}, \bibinfo {author} {\bibfnamefont {D.}~\bibnamefont
  {Bertho}}, \bibinfo {author} {\bibfnamefont {F.}~\bibnamefont {Kobbi}},
  \bibinfo {author} {\bibfnamefont {J.~L.}\ \bibnamefont {Robert}}, \bibinfo
  {author} {\bibfnamefont {G.~E.}\ \bibnamefont {Pikus}}, \bibinfo {author}
  {\bibfnamefont {F.~G.}\ \bibnamefont {Pikus}}, \bibinfo {author}
  {\bibfnamefont {S.~V.}\ \bibnamefont {Iordanskii}}, \bibinfo {author}
  {\bibfnamefont {V.}~\bibnamefont {Mosser}}, \bibinfo {author} {\bibfnamefont
  {K.}~\bibnamefont {Zekentes}}, \ and\ \bibinfo {author} {\bibfnamefont
  {Y.~B.}\ \bibnamefont {Lyanda-Geller}},\ }\href {\doibase
  10.1103/physrevb.53.3912} {\bibfield  {journal} {\bibinfo  {journal} {Phys.
  Rev. B}\ }\textbf {\bibinfo {volume} {53}},\ \bibinfo {pages} {3912}
  (\bibinfo {year} {1996})},\ \Eprint {http://arxiv.org/abs/9602068v1}
  {arXiv:9602068v1 [arXiv:cond-mat]} \BibitemShut {NoStop}%
\bibitem [{\citenamefont {Meyer}\ \emph {et~al.}(2002)\citenamefont {Meyer},
  \citenamefont {Fal'ko},\ and\ \citenamefont {Altshuler}}]{Meyer2002a}%
  \BibitemOpen
  \bibfield  {author} {\bibinfo {author} {\bibfnamefont {J.~S.}\ \bibnamefont
  {Meyer}}, \bibinfo {author} {\bibfnamefont {V.~I.}\ \bibnamefont {Fal'ko}}, \
  and\ \bibinfo {author} {\bibfnamefont {B.~L.}\ \bibnamefont {Altshuler}},\
  }in\ \href {\doibase 10.1007/978-94-010-0530-2_7} {\emph {\bibinfo
  {booktitle} {Strongly Correlated Fermions and Bosons in Low-Dimensional
  Disordered Systems}}}\ (\bibinfo  {publisher} {Springer Netherlands},\
  \bibinfo {year} {2002})\ pp.\ \bibinfo {pages} {117--164},\ \Eprint
  {http://arxiv.org/abs/0206024} {arXiv:0206024 [cond-mat]} \BibitemShut
  {NoStop}%
\bibitem [{\citenamefont {Wenk}\ and\ \citenamefont
  {Kettemann}(2010)}]{Wenk10}%
  \BibitemOpen
  \bibfield  {author} {\bibinfo {author} {\bibfnamefont {P.}~\bibnamefont
  {Wenk}}\ and\ \bibinfo {author} {\bibfnamefont {S.}~\bibnamefont
  {Kettemann}},\ }\href {\doibase 10.1103/PhysRevB.81.125309} {\bibfield
  {journal} {\bibinfo  {journal} {Phys. Rev. B}\ }\textbf {\bibinfo {volume}
  {81}},\ \bibinfo {pages} {125309} (\bibinfo {year} {2010})}\BibitemShut
  {NoStop}%
\bibitem [{\citenamefont {Kammermeier}\ \emph {et~al.}(2016)\citenamefont
  {Kammermeier}, \citenamefont {Wenk}, \citenamefont {Schliemann},
  \citenamefont {Heedt},\ and\ \citenamefont {Sch\"apers}}]{Kammermeier16}%
  \BibitemOpen
  \bibfield  {author} {\bibinfo {author} {\bibfnamefont {M.}~\bibnamefont
  {Kammermeier}}, \bibinfo {author} {\bibfnamefont {P.}~\bibnamefont {Wenk}},
  \bibinfo {author} {\bibfnamefont {J.}~\bibnamefont {Schliemann}}, \bibinfo
  {author} {\bibfnamefont {S.}~\bibnamefont {Heedt}}, \ and\ \bibinfo {author}
  {\bibfnamefont {T.}~\bibnamefont {Sch\"apers}},\ }\href {\doibase
  10.1103/PhysRevB.93.205306} {\bibfield  {journal} {\bibinfo  {journal} {Phys.
  Rev. B}\ }\textbf {\bibinfo {volume} {93}},\ \bibinfo {pages} {205306}
  (\bibinfo {year} {2016})}\BibitemShut {NoStop}%
\bibitem [{\citenamefont {Kettemann}(2007)}]{Kettemann07}%
  \BibitemOpen
  \bibfield  {author} {\bibinfo {author} {\bibfnamefont {S.}~\bibnamefont
  {Kettemann}},\ }\href {\doibase 10.1103/PhysRevLett.98.176808} {\bibfield
  {journal} {\bibinfo  {journal} {Phys. Rev. Lett.}\ }\textbf {\bibinfo
  {volume} {98}},\ \bibinfo {pages} {176808} (\bibinfo {year}
  {2007})}\BibitemShut {NoStop}%
\bibitem [{\citenamefont {Grifoni}\ and\ \citenamefont
  {H\"anggi}(1998)}]{Grifoni98}%
  \BibitemOpen
  \bibfield  {author} {\bibinfo {author} {\bibfnamefont {M.}~\bibnamefont
  {Grifoni}}\ and\ \bibinfo {author} {\bibfnamefont {P.}~\bibnamefont
  {H\"anggi}},\ }\href {\doibase https://doi.org/10.1016/S0370-1573(98)00022-2}
  {\bibfield  {journal} {\bibinfo  {journal} {Physics Reports}\ }\textbf
  {\bibinfo {volume} {304}},\ \bibinfo {pages} {229} (\bibinfo {year}
  {1998})}\BibitemShut {NoStop}%
\bibitem [{\citenamefont {Kitagawa}\ \emph {et~al.}(2011)\citenamefont
  {Kitagawa}, \citenamefont {Oka}, \citenamefont {Brataas}, \citenamefont
  {Fu},\ and\ \citenamefont {Demler}}]{Kitagawa11}%
  \BibitemOpen
  \bibfield  {author} {\bibinfo {author} {\bibfnamefont {T.}~\bibnamefont
  {Kitagawa}}, \bibinfo {author} {\bibfnamefont {T.}~\bibnamefont {Oka}},
  \bibinfo {author} {\bibfnamefont {A.}~\bibnamefont {Brataas}}, \bibinfo
  {author} {\bibfnamefont {L.}~\bibnamefont {Fu}}, \ and\ \bibinfo {author}
  {\bibfnamefont {E.}~\bibnamefont {Demler}},\ }\href {\doibase
  10.1103/PhysRevB.84.235108} {\bibfield  {journal} {\bibinfo  {journal} {Phys.
  Rev. B}\ }\textbf {\bibinfo {volume} {84}},\ \bibinfo {pages} {235108}
  (\bibinfo {year} {2011})}\BibitemShut {NoStop}%
\bibitem [{\citenamefont {Zhou}\ and\ \citenamefont {Wu}(2011)}]{Zhou11}%
  \BibitemOpen
  \bibfield  {author} {\bibinfo {author} {\bibfnamefont {Y.}~\bibnamefont
  {Zhou}}\ and\ \bibinfo {author} {\bibfnamefont {M.~W.}\ \bibnamefont {Wu}},\
  }\href {\doibase 10.1103/PhysRevB.83.245436} {\bibfield  {journal} {\bibinfo
  {journal} {Phys. Rev. B}\ }\textbf {\bibinfo {volume} {83}},\ \bibinfo
  {pages} {245436} (\bibinfo {year} {2011})}\BibitemShut {NoStop}%
\bibitem [{\citenamefont {Oka}\ and\ \citenamefont {Aoki}(2009)}]{Oka}%
  \BibitemOpen
  \bibfield  {author} {\bibinfo {author} {\bibfnamefont {T.}~\bibnamefont
  {Oka}}\ and\ \bibinfo {author} {\bibfnamefont {H.}~\bibnamefont {Aoki}},\
  }\href {\doibase 10.1103/PhysRevB.79.081406} {\bibfield  {journal} {\bibinfo
  {journal} {Phys. Rev. B}\ }\textbf {\bibinfo {volume} {79}},\ \bibinfo
  {pages} {081406(R)} (\bibinfo {year} {2009})}\BibitemShut {NoStop}%
\bibitem [{\citenamefont {Morina}\ \emph {et~al.}(2015)\citenamefont {Morina},
  \citenamefont {Kibis}, \citenamefont {Pervishko},\ and\ \citenamefont
  {Shelykh}}]{Morina15}%
  \BibitemOpen
  \bibfield  {author} {\bibinfo {author} {\bibfnamefont {S.}~\bibnamefont
  {Morina}}, \bibinfo {author} {\bibfnamefont {O.~V.}\ \bibnamefont {Kibis}},
  \bibinfo {author} {\bibfnamefont {A.~A.}\ \bibnamefont {Pervishko}}, \ and\
  \bibinfo {author} {\bibfnamefont {I.~A.}\ \bibnamefont {Shelykh}},\ }\href
  {\doibase 10.1103/PhysRevB.91.155312} {\bibfield  {journal} {\bibinfo
  {journal} {Phys. Rev. B}\ }\textbf {\bibinfo {volume} {91}},\ \bibinfo
  {pages} {155312} (\bibinfo {year} {2015})}\BibitemShut {NoStop}%
\bibitem [{\citenamefont {Skvortsov}(1998)}]{Skvortsov1998}%
  \BibitemOpen
  \bibfield  {author} {\bibinfo {author} {\bibfnamefont {M.~A.}\ \bibnamefont
  {Skvortsov}},\ }\href {\doibase 10.1134/1.567636} {\bibfield  {journal}
  {\bibinfo  {journal} {Journal of Experimental and Theoretical Physics
  Letters}\ }\textbf {\bibinfo {volume} {67}},\ \bibinfo {pages} {133}
  (\bibinfo {year} {1998})}\BibitemShut {NoStop}%
\bibitem [{\citenamefont {Pervishko}\ \emph {et~al.}(2015)\citenamefont
  {Pervishko}, \citenamefont {Kibis}, \citenamefont {Morina},\ and\
  \citenamefont {Shelykh}}]{Shelykh15}%
  \BibitemOpen
  \bibfield  {author} {\bibinfo {author} {\bibfnamefont {A.~A.}\ \bibnamefont
  {Pervishko}}, \bibinfo {author} {\bibfnamefont {O.~V.}\ \bibnamefont
  {Kibis}}, \bibinfo {author} {\bibfnamefont {S.}~\bibnamefont {Morina}}, \
  and\ \bibinfo {author} {\bibfnamefont {I.~A.}\ \bibnamefont {Shelykh}},\
  }\href {\doibase 10.1103/PhysRevB.92.205403} {\bibfield  {journal} {\bibinfo
  {journal} {Phys. Rev. B}\ }\textbf {\bibinfo {volume} {92}},\ \bibinfo
  {pages} {205403} (\bibinfo {year} {2015})}\BibitemShut {NoStop}%
\bibitem [{\citenamefont {Shi}\ and\ \citenamefont {Xie}(2003)}]{Shi03}%
  \BibitemOpen
  \bibfield  {author} {\bibinfo {author} {\bibfnamefont {J.}~\bibnamefont
  {Shi}}\ and\ \bibinfo {author} {\bibfnamefont {X.~C.}\ \bibnamefont {Xie}},\
  }\href {\doibase 10.1103/PhysRevLett.91.086801} {\bibfield  {journal}
  {\bibinfo  {journal} {Phys. Rev. Lett.}\ }\textbf {\bibinfo {volume} {91}},\
  \bibinfo {pages} {086801} (\bibinfo {year} {2003})}\BibitemShut {NoStop}%
\bibitem [{\citenamefont {Dehghani}\ \emph {et~al.}(2015)\citenamefont
  {Dehghani}, \citenamefont {Oka},\ and\ \citenamefont {Mitra}}]{Dehghani15}%
  \BibitemOpen
  \bibfield  {author} {\bibinfo {author} {\bibfnamefont {H.}~\bibnamefont
  {Dehghani}}, \bibinfo {author} {\bibfnamefont {T.}~\bibnamefont {Oka}}, \
  and\ \bibinfo {author} {\bibfnamefont {A.}~\bibnamefont {Mitra}},\ }\href
  {\doibase 10.1103/PhysRevB.91.155422} {\bibfield  {journal} {\bibinfo
  {journal} {Phys. Rev. B}\ }\textbf {\bibinfo {volume} {91}},\ \bibinfo
  {pages} {155422} (\bibinfo {year} {2015})}\BibitemShut {NoStop}%
\bibitem [{\citenamefont {Kibis}(2014)}]{Kibis14}%
  \BibitemOpen
  \bibfield  {author} {\bibinfo {author} {\bibfnamefont {O.~V.}\ \bibnamefont
  {Kibis}},\ }\href {\doibase https://doi.org/10.1209/0295-5075/107/57003}
  {\bibfield  {journal} {\bibinfo  {journal} {EPL (Europhysics Letters)}\
  }\textbf {\bibinfo {volume} {107}},\ \bibinfo {pages} {57003} (\bibinfo
  {year} {2014})}\BibitemShut {NoStop}%
\bibitem [{\citenamefont {Herrmann}\ \emph {et~al.}(2017)\citenamefont
  {Herrmann}, \citenamefont {Kvon}, \citenamefont {Dmitriev}, \citenamefont
  {Kozlov}, \citenamefont {Jentzsch}, \citenamefont {Schneider}, \citenamefont
  {Schell}, \citenamefont {Bel'kov}, \citenamefont {Bayer}, \citenamefont
  {Schuh}, \citenamefont {Bougeard}, \citenamefont {Kuczmik}, \citenamefont
  {Oltscher}, \citenamefont {Weiss},\ and\ \citenamefont
  {Ganichev}}]{Ganichev17}%
  \BibitemOpen
  \bibfield  {author} {\bibinfo {author} {\bibfnamefont {T.}~\bibnamefont
  {Herrmann}}, \bibinfo {author} {\bibfnamefont {Z.~D.}\ \bibnamefont {Kvon}},
  \bibinfo {author} {\bibfnamefont {I.~A.}\ \bibnamefont {Dmitriev}}, \bibinfo
  {author} {\bibfnamefont {D.~A.}\ \bibnamefont {Kozlov}}, \bibinfo {author}
  {\bibfnamefont {B.}~\bibnamefont {Jentzsch}}, \bibinfo {author}
  {\bibfnamefont {M.}~\bibnamefont {Schneider}}, \bibinfo {author}
  {\bibfnamefont {L.}~\bibnamefont {Schell}}, \bibinfo {author} {\bibfnamefont
  {V.~V.}\ \bibnamefont {Bel'kov}}, \bibinfo {author} {\bibfnamefont
  {A.}~\bibnamefont {Bayer}}, \bibinfo {author} {\bibfnamefont
  {D.}~\bibnamefont {Schuh}}, \bibinfo {author} {\bibfnamefont
  {D.}~\bibnamefont {Bougeard}}, \bibinfo {author} {\bibfnamefont
  {T.}~\bibnamefont {Kuczmik}}, \bibinfo {author} {\bibfnamefont
  {M.}~\bibnamefont {Oltscher}}, \bibinfo {author} {\bibfnamefont
  {D.}~\bibnamefont {Weiss}}, \ and\ \bibinfo {author} {\bibfnamefont {S.~D.}\
  \bibnamefont {Ganichev}},\ }\href {\doibase 10.1103/PhysRevB.96.115449}
  {\bibfield  {journal} {\bibinfo  {journal} {Phys. Rev. B}\ }\textbf {\bibinfo
  {volume} {96}},\ \bibinfo {pages} {115449} (\bibinfo {year}
  {2017})}\BibitemShut {NoStop}%
\bibitem [{\citenamefont {Genske}\ and\ \citenamefont
  {Rosch}(2015)}]{Genske15}%
  \BibitemOpen
  \bibfield  {author} {\bibinfo {author} {\bibfnamefont {M.}~\bibnamefont
  {Genske}}\ and\ \bibinfo {author} {\bibfnamefont {A.}~\bibnamefont {Rosch}},\
  }\href {\doibase 10.1103/PhysRevA.92.062108} {\bibfield  {journal} {\bibinfo
  {journal} {Phys. Rev. A}\ }\textbf {\bibinfo {volume} {92}},\ \bibinfo
  {pages} {062108} (\bibinfo {year} {2015})}\BibitemShut {NoStop}%
\bibitem [{\citenamefont {Esin}\ \emph {et~al.}(2018)\citenamefont {Esin},
  \citenamefont {Rudner}, \citenamefont {Refael},\ and\ \citenamefont
  {Lindner}}]{Esin2018}%
  \BibitemOpen
  \bibfield  {author} {\bibinfo {author} {\bibfnamefont {I.}~\bibnamefont
  {Esin}}, \bibinfo {author} {\bibfnamefont {M.~S.}\ \bibnamefont {Rudner}},
  \bibinfo {author} {\bibfnamefont {G.}~\bibnamefont {Refael}}, \ and\ \bibinfo
  {author} {\bibfnamefont {N.~H.}\ \bibnamefont {Lindner}},\ }\href {\doibase
  10.1103/physrevb.97.245401} {\bibfield  {journal} {\bibinfo  {journal}
  {Physical Review B}\ }\textbf {\bibinfo {volume} {97}} (\bibinfo {year}
  {2018}),\ 10.1103/physrevb.97.245401}\BibitemShut {NoStop}%
\bibitem [{\citenamefont {Titum}\ \emph {et~al.}(2015)\citenamefont {Titum},
  \citenamefont {Lindner}, \citenamefont {Rechtsman},\ and\ \citenamefont
  {Refael}}]{Paraj15}%
  \BibitemOpen
  \bibfield  {author} {\bibinfo {author} {\bibfnamefont {P.}~\bibnamefont
  {Titum}}, \bibinfo {author} {\bibfnamefont {N.~H.}\ \bibnamefont {Lindner}},
  \bibinfo {author} {\bibfnamefont {M.~C.}\ \bibnamefont {Rechtsman}}, \ and\
  \bibinfo {author} {\bibfnamefont {G.}~\bibnamefont {Refael}},\ }\href
  {\doibase 10.1103/PhysRevLett.114.056801} {\bibfield  {journal} {\bibinfo
  {journal} {Phys. Rev. Lett.}\ }\textbf {\bibinfo {volume} {114}},\ \bibinfo
  {pages} {056801} (\bibinfo {year} {2015})}\BibitemShut {NoStop}%
\bibitem [{\citenamefont {Titum}\ \emph {et~al.}(2017)\citenamefont {Titum},
  \citenamefont {Lindner},\ and\ \citenamefont {Refael}}]{Paraj17}%
  \BibitemOpen
  \bibfield  {author} {\bibinfo {author} {\bibfnamefont {P.}~\bibnamefont
  {Titum}}, \bibinfo {author} {\bibfnamefont {N.~H.}\ \bibnamefont {Lindner}},
  \ and\ \bibinfo {author} {\bibfnamefont {G.}~\bibnamefont {Refael}},\ }\href
  {\doibase 10.1103/PhysRevB.96.054207} {\bibfield  {journal} {\bibinfo
  {journal} {Phys. Rev. B}\ }\textbf {\bibinfo {volume} {96}},\ \bibinfo
  {pages} {054207} (\bibinfo {year} {2017})}\BibitemShut {NoStop}%
\bibitem [{\citenamefont {Titum}(2016)}]{ParajPhd}%
  \BibitemOpen
  \bibfield  {author} {\bibinfo {author} {\bibfnamefont {P.}~\bibnamefont
  {Titum}},\ }{\selectlanguage {English}\emph {\bibinfo {title} {{Disorder
  Driven Transitions in Non-Equilibrium Quantum Systems}}}},\ \href {\doibase
  10.7907/Z9MK69VV} {Ph.D. thesis},\ \bibinfo  {school} {California Institute
  of Technology} (\bibinfo {year} {2016})\BibitemShut {NoStop}%
\bibitem [{\citenamefont {Floquet}(1883)}]{Floquet83}%
  \BibitemOpen
  \bibfield  {author} {\bibinfo {author} {\bibfnamefont {G.}~\bibnamefont
  {Floquet}},\ }\href {\doibase 10.24033/asens.220} {\bibfield  {journal}
  {\bibinfo  {journal} {Annales scientifiques de l'\'Ecole Normale
  Sup\'erieure}\ }\textbf {\bibinfo {volume} {12}},\ \bibinfo {pages} {47}
  (\bibinfo {year} {1883})}\BibitemShut {NoStop}%
\bibitem [{\citenamefont {Holthaus}(2015)}]{Holthaus15}%
  \BibitemOpen
  \bibfield  {author} {\bibinfo {author} {\bibfnamefont {M.}~\bibnamefont
  {Holthaus}},\ }\href {\doibase https://doi.org/10.1088/0953-4075/49/1/013001}
  {\bibfield  {journal} {\bibinfo  {journal} {Journal of Physics B: Atomic,
  Molecular and Optical Physics}\ }\textbf {\bibinfo {volume} {49}},\ \bibinfo
  {pages} {013001} (\bibinfo {year} {2015})}\BibitemShut {NoStop}%
\bibitem [{\citenamefont {Zel'Dovich}(1967)}]{zel1967quasienergy}%
  \BibitemOpen
  \bibfield  {author} {\bibinfo {author} {\bibfnamefont {Y.~B.}\ \bibnamefont
  {Zel'Dovich}},\ }\href@noop {} {\bibfield  {journal} {\bibinfo  {journal}
  {Soviet Physics JETP}\ }\textbf {\bibinfo {volume} {24}},\ \bibinfo {pages}
  {1006} (\bibinfo {year} {1967})}\BibitemShut {NoStop}%
\bibitem [{\citenamefont {Ritus}(1967)}]{ritus1967shift}%
  \BibitemOpen
  \bibfield  {author} {\bibinfo {author} {\bibfnamefont {V.}~\bibnamefont
  {Ritus}},\ }\href@noop {} {\bibfield  {journal} {\bibinfo  {journal} {Sov.
  Phys. JETP}\ }\textbf {\bibinfo {volume} {24}},\ \bibinfo {pages} {1041}
  (\bibinfo {year} {1967})}\BibitemShut {NoStop}%
\bibitem [{\citenamefont {Mahan}(2013)}]{Mahan2013}%
  \BibitemOpen
  \bibfield  {author} {\bibinfo {author} {\bibfnamefont {G.~D.}\ \bibnamefont
  {Mahan}},\ }\href@noop {} {\emph {\bibinfo {title} {Many-particle physics}}}\
  (\bibinfo  {publisher} {Springer Science \& Business Media},\ \bibinfo {year}
  {2013})\BibitemShut {NoStop}%
\bibitem [{\citenamefont {Gupta}\ \emph {et~al.}(2003)\citenamefont {Gupta},
  \citenamefont {Alon},\ and\ \citenamefont {Moiseyev}}]{Gupta03}%
  \BibitemOpen
  \bibfield  {author} {\bibinfo {author} {\bibfnamefont {A.~K.}\ \bibnamefont
  {Gupta}}, \bibinfo {author} {\bibfnamefont {O.~E.}\ \bibnamefont {Alon}}, \
  and\ \bibinfo {author} {\bibfnamefont {N.}~\bibnamefont {Moiseyev}},\ }\href
  {\doibase 10.1103/PhysRevB.68.205101} {\bibfield  {journal} {\bibinfo
  {journal} {Phys. Rev. B}\ }\textbf {\bibinfo {volume} {68}},\ \bibinfo
  {pages} {205101} (\bibinfo {year} {2003})}\BibitemShut {NoStop}%
\bibitem [{\citenamefont {Hsu}\ and\ \citenamefont {Reichl}(2006)}]{Hsu06}%
  \BibitemOpen
  \bibfield  {author} {\bibinfo {author} {\bibfnamefont {H.}~\bibnamefont
  {Hsu}}\ and\ \bibinfo {author} {\bibfnamefont {L.~E.}\ \bibnamefont
  {Reichl}},\ }\href {\doibase 10.1103/PhysRevB.74.115406} {\bibfield
  {journal} {\bibinfo  {journal} {Phys. Rev. B}\ }\textbf {\bibinfo {volume}
  {74}},\ \bibinfo {pages} {115406} (\bibinfo {year} {2006})}\BibitemShut
  {NoStop}%
\bibitem [{\citenamefont {Scholz}\ \emph {et~al.}(2013)\citenamefont {Scholz},
  \citenamefont {L\'opez},\ and\ \citenamefont {Schliemann}}]{Scholz13}%
  \BibitemOpen
  \bibfield  {author} {\bibinfo {author} {\bibfnamefont {A.}~\bibnamefont
  {Scholz}}, \bibinfo {author} {\bibfnamefont {A.}~\bibnamefont {L\'opez}}, \
  and\ \bibinfo {author} {\bibfnamefont {J.}~\bibnamefont {Schliemann}},\
  }\href {\doibase 10.1103/PhysRevB.88.045118} {\bibfield  {journal} {\bibinfo
  {journal} {Phys. Rev. B}\ }\textbf {\bibinfo {volume} {88}},\ \bibinfo
  {pages} {045118} (\bibinfo {year} {2013})}\BibitemShut {NoStop}%
\bibitem [{\citenamefont {Shirai}\ \emph {et~al.}(2015)\citenamefont {Shirai},
  \citenamefont {Mori},\ and\ \citenamefont {Miyashita}}]{Shirai15}%
  \BibitemOpen
  \bibfield  {author} {\bibinfo {author} {\bibfnamefont {T.}~\bibnamefont
  {Shirai}}, \bibinfo {author} {\bibfnamefont {T.}~\bibnamefont {Mori}}, \ and\
  \bibinfo {author} {\bibfnamefont {S.}~\bibnamefont {Miyashita}},\ }\href
  {\doibase 10.1103/PhysRevE.91.030101} {\bibfield  {journal} {\bibinfo
  {journal} {Phys. Rev. E}\ }\textbf {\bibinfo {volume} {91}},\ \bibinfo
  {pages} {030101(R)} (\bibinfo {year} {2015})}\BibitemShut {NoStop}%
\bibitem [{\citenamefont {Shirai}\ \emph {et~al.}(2016)\citenamefont {Shirai},
  \citenamefont {Thingna}, \citenamefont {Mori}, \citenamefont {Denisov},
  \citenamefont {Hänggi},\ and\ \citenamefont {Miyashita}}]{Shirai16}%
  \BibitemOpen
  \bibfield  {author} {\bibinfo {author} {\bibfnamefont {T.}~\bibnamefont
  {Shirai}}, \bibinfo {author} {\bibfnamefont {J.}~\bibnamefont {Thingna}},
  \bibinfo {author} {\bibfnamefont {T.}~\bibnamefont {Mori}}, \bibinfo {author}
  {\bibfnamefont {S.}~\bibnamefont {Denisov}}, \bibinfo {author} {\bibfnamefont
  {P.}~\bibnamefont {Hänggi}}, \ and\ \bibinfo {author} {\bibfnamefont
  {S.}~\bibnamefont {Miyashita}},\ }\href {\doibase
  10.1088/1367-2630/18/5/053008} {\bibfield  {journal} {\bibinfo  {journal}
  {New Journal of Physics}\ }\textbf {\bibinfo {volume} {18}},\ \bibinfo
  {pages} {053008} (\bibinfo {year} {2016})}\BibitemShut {NoStop}%
\bibitem [{\citenamefont {Bilitewski}\ and\ \citenamefont
  {Cooper}(2015)}]{Bilitewski15}%
  \BibitemOpen
  \bibfield  {author} {\bibinfo {author} {\bibfnamefont {T.}~\bibnamefont
  {Bilitewski}}\ and\ \bibinfo {author} {\bibfnamefont {N.~R.}\ \bibnamefont
  {Cooper}},\ }\href {\doibase 10.1103/PhysRevA.91.033601} {\bibfield
  {journal} {\bibinfo  {journal} {Phys. Rev. A}\ }\textbf {\bibinfo {volume}
  {91}},\ \bibinfo {pages} {033601} (\bibinfo {year} {2015})}\BibitemShut
  {NoStop}%
\bibitem [{\citenamefont {Martinez}(2003)}]{Martinez03}%
  \BibitemOpen
  \bibfield  {author} {\bibinfo {author} {\bibfnamefont {D.~F.}\ \bibnamefont
  {Martinez}},\ }\href {\doibase https://doi.org/10.1088/0305-4470/36/38/302}
  {\bibfield  {journal} {\bibinfo  {journal} {Journal of Physics A:
  Mathematical and General}\ }\textbf {\bibinfo {volume} {36}},\ \bibinfo
  {pages} {9827} (\bibinfo {year} {2003})}\BibitemShut {NoStop}%
\bibitem [{\citenamefont {Drese}\ and\ \citenamefont
  {Holthaus}(1999)}]{Drese99}%
  \BibitemOpen
  \bibfield  {author} {\bibinfo {author} {\bibfnamefont {K.}~\bibnamefont
  {Drese}}\ and\ \bibinfo {author} {\bibfnamefont {M.}~\bibnamefont
  {Holthaus}},\ }\href {\doibase 10.1007/s100530050236} {\bibfield  {journal}
  {\bibinfo  {journal} {The European Physical Journal D - Atomic, Molecular,
  Optical and Plasma Physics}\ }\textbf {\bibinfo {volume} {5}},\ \bibinfo
  {pages} {119} (\bibinfo {year} {1999})}\BibitemShut {NoStop}%
\bibitem [{\citenamefont {Sambe}(1973)}]{Sambe73}%
  \BibitemOpen
  \bibfield  {author} {\bibinfo {author} {\bibfnamefont {H.}~\bibnamefont
  {Sambe}},\ }\href {\doibase 10.1103/PhysRevA.7.2203} {\bibfield  {journal}
  {\bibinfo  {journal} {Phys. Rev. A}\ }\textbf {\bibinfo {volume} {7}},\
  \bibinfo {pages} {2203} (\bibinfo {year} {1973})}\BibitemShut {NoStop}%
\bibitem [{\citenamefont {Martinez}(2005)}]{Martinez05}%
  \BibitemOpen
  \bibfield  {author} {\bibinfo {author} {\bibfnamefont {D.~F.}\ \bibnamefont
  {Martinez}},\ }\href {\doibase https://doi.org/10.1088/0305-4470/38/46/006}
  {\bibfield  {journal} {\bibinfo  {journal} {Journal of Physics A:
  Mathematical and General}\ }\textbf {\bibinfo {volume} {38}},\ \bibinfo
  {pages} {9979} (\bibinfo {year} {2005})}\BibitemShut {NoStop}%
\bibitem [{\citenamefont {Martinez}\ and\ \citenamefont
  {Molina}(2006)}]{Martinez06}%
  \BibitemOpen
  \bibfield  {author} {\bibinfo {author} {\bibfnamefont {D.~F.}\ \bibnamefont
  {Martinez}}\ and\ \bibinfo {author} {\bibfnamefont {R.~A.}\ \bibnamefont
  {Molina}},\ }\href {\doibase 10.1140/epjb/e2006-00293-7} {\bibfield
  {journal} {\bibinfo  {journal} {The European Physical Journal B}\ }\textbf
  {\bibinfo {volume} {52}},\ \bibinfo {pages} {281} (\bibinfo {year}
  {2006})}\BibitemShut {NoStop}%
\bibitem [{\citenamefont {Eissing}\ \emph
  {et~al.}(2016{\natexlab{a}})\citenamefont {Eissing}, \citenamefont {Meden},\
  and\ \citenamefont {Kennes}}]{Eissing16}%
  \BibitemOpen
  \bibfield  {author} {\bibinfo {author} {\bibfnamefont {A.~K.}\ \bibnamefont
  {Eissing}}, \bibinfo {author} {\bibfnamefont {V.}~\bibnamefont {Meden}}, \
  and\ \bibinfo {author} {\bibfnamefont {D.~M.}\ \bibnamefont {Kennes}},\
  }\href {\doibase 10.1103/PhysRevLett.116.026801} {\bibfield  {journal}
  {\bibinfo  {journal} {Phys. Rev. Lett.}\ }\textbf {\bibinfo {volume} {116}},\
  \bibinfo {pages} {026801} (\bibinfo {year} {2016}{\natexlab{a}})}\BibitemShut
  {NoStop}%
\bibitem [{\citenamefont {Eissing}\ \emph
  {et~al.}(2016{\natexlab{b}})\citenamefont {Eissing}, \citenamefont {Meden},\
  and\ \citenamefont {Kennes}}]{Eissing16B}%
  \BibitemOpen
  \bibfield  {author} {\bibinfo {author} {\bibfnamefont {A.~K.}\ \bibnamefont
  {Eissing}}, \bibinfo {author} {\bibfnamefont {V.}~\bibnamefont {Meden}}, \
  and\ \bibinfo {author} {\bibfnamefont {D.~M.}\ \bibnamefont {Kennes}},\
  }\href {\doibase 10.1103/PhysRevB.94.245116} {\bibfield  {journal} {\bibinfo
  {journal} {Phys. Rev. B}\ }\textbf {\bibinfo {volume} {94}},\ \bibinfo
  {pages} {245116} (\bibinfo {year} {2016}{\natexlab{b}})}\BibitemShut
  {NoStop}%
\bibitem [{\citenamefont {Taberner}(2017)}]{Taberner17}%
  \BibitemOpen
  \bibfield  {author} {\bibinfo {author} {\bibfnamefont {C.~O.}\ \bibnamefont
  {Taberner}},\ }\emph {\bibinfo {title} {Periodically driven S-QD-S
  junctions}},\ \href@noop {} {Master's thesis},\ \bibinfo  {school}
  {University of Copenhagen} (\bibinfo {year} {2017})\BibitemShut {NoStop}%
\bibitem [{\citenamefont {Aoki}\ \emph {et~al.}(2014)\citenamefont {Aoki},
  \citenamefont {Tsuji}, \citenamefont {Eckstein}, \citenamefont {Kollar},
  \citenamefont {Oka},\ and\ \citenamefont {Werner}}]{Aoki14}%
  \BibitemOpen
  \bibfield  {author} {\bibinfo {author} {\bibfnamefont {H.}~\bibnamefont
  {Aoki}}, \bibinfo {author} {\bibfnamefont {N.}~\bibnamefont {Tsuji}},
  \bibinfo {author} {\bibfnamefont {M.}~\bibnamefont {Eckstein}}, \bibinfo
  {author} {\bibfnamefont {M.}~\bibnamefont {Kollar}}, \bibinfo {author}
  {\bibfnamefont {T.}~\bibnamefont {Oka}}, \ and\ \bibinfo {author}
  {\bibfnamefont {P.}~\bibnamefont {Werner}},\ }\href {\doibase
  10.1103/RevModPhys.86.779} {\bibfield  {journal} {\bibinfo  {journal} {Rev.
  Mod. Phys.}\ }\textbf {\bibinfo {volume} {86}},\ \bibinfo {pages} {779}
  (\bibinfo {year} {2014})}\BibitemShut {NoStop}%
\bibitem [{\citenamefont {Rentrop}\ \emph {et~al.}(2014)\citenamefont
  {Rentrop}, \citenamefont {Jakobs},\ and\ \citenamefont {Meden}}]{Rentrop14}%
  \BibitemOpen
  \bibfield  {author} {\bibinfo {author} {\bibfnamefont {J.~F.}\ \bibnamefont
  {Rentrop}}, \bibinfo {author} {\bibfnamefont {S.~G.}\ \bibnamefont {Jakobs}},
  \ and\ \bibinfo {author} {\bibfnamefont {V.}~\bibnamefont {Meden}},\ }\href
  {\doibase 10.1103/PhysRevB.89.235110} {\bibfield  {journal} {\bibinfo
  {journal} {Phys. Rev. B}\ }\textbf {\bibinfo {volume} {89}},\ \bibinfo
  {pages} {235110} (\bibinfo {year} {2014})}\BibitemShut {NoStop}%
\bibitem [{\citenamefont {Wu}\ and\ \citenamefont {Cao}(2008)}]{Wu08}%
  \BibitemOpen
  \bibfield  {author} {\bibinfo {author} {\bibfnamefont {B.~H.}\ \bibnamefont
  {Wu}}\ and\ \bibinfo {author} {\bibfnamefont {J.~C.}\ \bibnamefont {Cao}},\
  }\href {\doibase 10.1088/0953-8984/20/8/085224} {\bibfield  {journal}
  {\bibinfo  {journal} {Journal of Physics: Condensed Matter}\ }\textbf
  {\bibinfo {volume} {20}},\ \bibinfo {pages} {085224} (\bibinfo {year}
  {2008})}\BibitemShut {NoStop}%
\bibitem [{\citenamefont {Shirley}(1963)}]{Shirley63}%
  \BibitemOpen
  \bibfield  {author} {\bibinfo {author} {\bibfnamefont {J.~H.}\ \bibnamefont
  {Shirley}},\ }\emph {\bibinfo {title} {Interaction of a quantum system with a
  strong oscillating field}},\ \href
  {http://resolver.caltech.edu/CaltechETD:etd-05142008-103758} {Ph.D. thesis},\
  \bibinfo  {school} {California Institute of Technology} (\bibinfo {year}
  {1963})\BibitemShut {NoStop}%
\bibitem [{\citenamefont {Shirley}(1965)}]{Shirley65}%
  \BibitemOpen
  \bibfield  {author} {\bibinfo {author} {\bibfnamefont {J.~H.}\ \bibnamefont
  {Shirley}},\ }\href {\doibase 10.1103/PhysRev.138.B979} {\bibfield  {journal}
  {\bibinfo  {journal} {Phys. Rev.}\ }\textbf {\bibinfo {volume} {138}},\
  \bibinfo {pages} {B979} (\bibinfo {year} {1965})}\BibitemShut {NoStop}%
\bibitem [{\citenamefont {Tsuji}\ \emph {et~al.}(2008)\citenamefont {Tsuji},
  \citenamefont {Oka},\ and\ \citenamefont {Aoki}}]{Tsuji08}%
  \BibitemOpen
  \bibfield  {author} {\bibinfo {author} {\bibfnamefont {N.}~\bibnamefont
  {Tsuji}}, \bibinfo {author} {\bibfnamefont {T.}~\bibnamefont {Oka}}, \ and\
  \bibinfo {author} {\bibfnamefont {H.}~\bibnamefont {Aoki}},\ }\href {\doibase
  10.1103/PhysRevB.78.235124} {\bibfield  {journal} {\bibinfo  {journal} {Phys.
  Rev. B}\ }\textbf {\bibinfo {volume} {78}},\ \bibinfo {pages} {235124}
  (\bibinfo {year} {2008})}\BibitemShut {NoStop}%
\bibitem [{\citenamefont {Henrik~Bruus}(2004)}]{BruusFlensberg}%
  \BibitemOpen
  \bibfield  {author} {\bibinfo {author} {\bibfnamefont {K.~F.}\ \bibnamefont
  {Henrik~Bruus}},\ }\href
  {https://www.ebook.de/de/product/23129589/henrik_bruus_karsten_flensberg_many_body_quantum_theory_in_condensed_matter_physics.html}
  {\emph {\bibinfo {title} {Many-Body Quantum Theory in Condensed Matter
  Physics}}}\ (\bibinfo  {publisher} {Oxford University Press},\ \bibinfo
  {address} {United States},\ \bibinfo {year} {2004})\BibitemShut {NoStop}%
\bibitem [{\citenamefont {Akkermans}\ and\ \citenamefont
  {Montambaux}(2007)}]{akkermans}%
  \BibitemOpen
  \bibfield  {author} {\bibinfo {author} {\bibfnamefont {E.}~\bibnamefont
  {Akkermans}}\ and\ \bibinfo {author} {\bibfnamefont {G.}~\bibnamefont
  {Montambaux}},\ }\href {\doibase 10.1017/CBO9780511618833} {\emph {\bibinfo
  {title} {Mesoscopic Physics of Electrons and Photons}}}\ (\bibinfo
  {publisher} {Cambridge University Press},\ \bibinfo {year}
  {2007})\BibitemShut {NoStop}%
\bibitem [{\citenamefont {Rammer}(2004)}]{rammer}%
  \BibitemOpen
  \bibfield  {author} {\bibinfo {author} {\bibfnamefont {J.}~\bibnamefont
  {Rammer}},\ }\href
  {https://www.amazon.com/Quantum-Transport-Theory-Frontiers-Physics/dp/0813342848?SubscriptionId=AKIAIOBINVZYXZQZ2U3A&tag=chimbori05-20&linkCode=xm2&camp=2025&creative=165953&creativeASIN=0813342848}
  {\emph {\bibinfo {title} {Quantum Transport Theory (Frontiers in
  Physics)}}},\ Frontiers in Physics\ (\bibinfo  {publisher} {Westview Press},\
  \bibinfo {year} {2004})\BibitemShut {NoStop}%
\end{thebibliography}%


\def\url#1{}
\begin{thebibliography}{9}%
\makeatletter
\providecommand \@ifxundefined [1]{%
 \@ifx{#1\undefined}
}%
\providecommand \@ifnum [1]{%
 \ifnum #1\expandafter \@firstoftwo
 \else \expandafter \@secondoftwo
 \fi
}%
\providecommand \@ifx [1]{%
 \ifx #1\expandafter \@firstoftwo
 \else \expandafter \@secondoftwo
 \fi
}%
\providecommand \natexlab [1]{#1}%
\providecommand \enquote  [1]{``#1''}%
\providecommand \bibnamefont  [1]{#1}%
\providecommand \bibfnamefont [1]{#1}%
\providecommand \citenamefont [1]{#1}%
\providecommand \href@noop [0]{\@secondoftwo}%
\providecommand \href [0]{\begingroup \@sanitize@url \@href}%
\providecommand \@href[1]{\@@startlink{#1}\@@href}%
\providecommand \@@href[1]{\endgroup#1\@@endlink}%
\providecommand \@sanitize@url [0]{\catcode `\\12\catcode `\$12\catcode
  `\&12\catcode `\#12\catcode `\^12\catcode `\_12\catcode `\%12\relax}%
\providecommand \@@startlink[1]{}%
\providecommand \@@endlink[0]{}%
\providecommand \url  [0]{\begingroup\@sanitize@url \@url }%
\providecommand \@url [1]{\endgroup\@href {#1}{\urlprefix }}%
\providecommand \urlprefix  [0]{URL }%
\providecommand \Eprint [0]{\href }%
\providecommand \doibase [0]{http://dx.doi.org/}%
\providecommand \selectlanguage [0]{\@gobble}%
\providecommand \bibinfo  [0]{\@secondoftwo}%
\providecommand \bibfield  [0]{\@secondoftwo}%
\providecommand \translation [1]{[#1]}%
\providecommand \BibitemOpen [0]{}%
\providecommand \bibitemStop [0]{}%
\providecommand \bibitemNoStop [0]{.\EOS\space}%
\providecommand \EOS [0]{\spacefactor3000\relax}%
\providecommand \BibitemShut  [1]{\csname bibitem#1\endcsname}%
\let\auto@bib@innerbib\@empty
\bibitem [{\citenamefont {Sambe}(1973)}]{Sambe73}%
  \BibitemOpen
  \bibfield  {author} {\bibinfo {author} {\bibfnamefont {H.}~\bibnamefont
  {Sambe}},\ }\href {\doibase 10.1103/PhysRevA.7.2203} {\bibfield  {journal}
  {\bibinfo  {journal} {Phys. Rev. A}\ }\textbf {\bibinfo {volume} {7}},\
  \bibinfo {pages} {2203} (\bibinfo {year} {1973})}\BibitemShut {NoStop}%
\bibitem [{\citenamefont {Akkermans}\ and\ \citenamefont
  {Montambaux}(2007)}]{akkermans}%
  \BibitemOpen
  \bibfield  {author} {\bibinfo {author} {\bibfnamefont {E.}~\bibnamefont
  {Akkermans}}\ and\ \bibinfo {author} {\bibfnamefont {G.}~\bibnamefont
  {Montambaux}},\ }\href {\doibase 10.1017/CBO9780511618833} {\emph {\bibinfo
  {title} {Mesoscopic Physics of Electrons and Photons}}}\ (\bibinfo
  {publisher} {Cambridge University Press},\ \bibinfo {year}
  {2007})\BibitemShut {NoStop}%
\bibitem [{\citenamefont {Dirac}(1927)}]{Dirac27}%
  \BibitemOpen
  \bibfield  {author} {\bibinfo {author} {\bibfnamefont {P.~A.~M.}\
  \bibnamefont {Dirac}},\ }\href {\doibase 10.1098/rspa.1927.0039} {\bibfield
  {journal} {\bibinfo  {journal} {Proceedings of the Royal Society A:
  Mathematical, Physical and Engineering Sciences}\ }\textbf {\bibinfo {volume}
  {114}},\ \bibinfo {pages} {243} (\bibinfo {year} {1927})}\BibitemShut
  {NoStop}%
\bibitem [{\citenamefont {Fermi}(1974)}]{Fermi50}%
  \BibitemOpen
  \bibfield  {author} {\bibinfo {author} {\bibfnamefont {E.}~\bibnamefont
  {Fermi}},\ }\href {https://books.google.de/books?id=WQtkYCWTcicC} {\emph
  {\bibinfo {title} {Nuclear Physics: A Course Given by Enrico Fermi at the
  University of Chicago}}},\ Midway reprint\ (\bibinfo  {publisher} {University
  of Chicago Press},\ \bibinfo {year} {1974})\BibitemShut {NoStop}%
\bibitem [{\citenamefont {Mahan}(2013)}]{Mahan2013}%
  \BibitemOpen
  \bibfield  {author} {\bibinfo {author} {\bibfnamefont {G.~D.}\ \bibnamefont
  {Mahan}},\ }\href@noop {} {\emph {\bibinfo {title} {Many-particle physics}}}\
  (\bibinfo  {publisher} {Springer Science \& Business Media},\ \bibinfo {year}
  {2013})\BibitemShut {NoStop}%
\bibitem [{\citenamefont {Sakurai}\ and\ \citenamefont
  {Napolitano}(2011)}]{sakurai}%
  \BibitemOpen
  \bibfield  {author} {\bibinfo {author} {\bibfnamefont {J.}~\bibnamefont
  {Sakurai}}\ and\ \bibinfo {author} {\bibfnamefont {J.}~\bibnamefont
  {Napolitano}},\ }\href {https://books.google.de/books?id=N4I-AQAACAAJ} {\emph
  {\bibinfo {title} {Modern Quantum Mechanics}}},\ 物理学经典教材\
  (\bibinfo  {publisher} {Addison-Wesley},\ \bibinfo {year} {2011})\BibitemShut
  {NoStop}%
\bibitem [{\citenamefont {Kitagawa}\ \emph {et~al.}(2011)\citenamefont
  {Kitagawa}, \citenamefont {Oka}, \citenamefont {Brataas}, \citenamefont
  {Fu},\ and\ \citenamefont {Demler}}]{Kitagawa11}%
  \BibitemOpen
  \bibfield  {author} {\bibinfo {author} {\bibfnamefont {T.}~\bibnamefont
  {Kitagawa}}, \bibinfo {author} {\bibfnamefont {T.}~\bibnamefont {Oka}},
  \bibinfo {author} {\bibfnamefont {A.}~\bibnamefont {Brataas}}, \bibinfo
  {author} {\bibfnamefont {L.}~\bibnamefont {Fu}}, \ and\ \bibinfo {author}
  {\bibfnamefont {E.}~\bibnamefont {Demler}},\ }\href {\doibase
  10.1103/PhysRevB.84.235108} {\bibfield  {journal} {\bibinfo  {journal} {Phys.
  Rev. B}\ }\textbf {\bibinfo {volume} {84}},\ \bibinfo {pages} {235108}
  (\bibinfo {year} {2011})}\BibitemShut {NoStop}%
\bibitem [{\citenamefont {Bilitewski}\ and\ \citenamefont
  {Cooper}(2015)}]{Bilitewski15}%
  \BibitemOpen
  \bibfield  {author} {\bibinfo {author} {\bibfnamefont {T.}~\bibnamefont
  {Bilitewski}}\ and\ \bibinfo {author} {\bibfnamefont {N.~R.}\ \bibnamefont
  {Cooper}},\ }\href {\doibase 10.1103/PhysRevA.91.033601} {\bibfield
  {journal} {\bibinfo  {journal} {Phys. Rev. A}\ }\textbf {\bibinfo {volume}
  {91}},\ \bibinfo {pages} {033601} (\bibinfo {year} {2015})}\BibitemShut
  {NoStop}%
\bibitem [{\citenamefont {Kibis}(2014)}]{Kibis14}%
  \BibitemOpen
  \bibfield  {author} {\bibinfo {author} {\bibfnamefont {O.~V.}\ \bibnamefont
  {Kibis}},\ }\href {\doibase https://doi.org/10.1209/0295-5075/107/57003}
  {\bibfield  {journal} {\bibinfo  {journal} {EPL (Europhysics Letters)}\
  }\textbf {\bibinfo {volume} {107}},\ \bibinfo {pages} {57003} (\bibinfo
  {year} {2014})}\BibitemShut {NoStop}%
\end{thebibliography}%


\def\url#1{}
%
\end{document}


\parindent0pt
\newcommand{\ua}[1]{u_{\alpha}^{n_{#1}}}
\newcommand{\ub}[1]{u_{\beta}^{n_{#1}}}

\newcommand{\ea}{\varepsilon_{\alpha}}
\newcommand{\eb}{\varepsilon_{\beta}}
\newcommand{\ih}{\frac{1}{\hbar}}
\newcommand{\G}{\mathcal{G}}
\newcommand{\se}{\boldsymbol{\Sigma}}
\newcommand{\kx}{k_{x}}
\newcommand{\ky}{k_{y}}
\newcommand{\gellmann}[1]{\boldsymbol{\lambda_{#1}}}
\newcommand{\Vimp}{\mathcal{V}_{\mathrm{imp}}}

\newcommand{\f}[1]{\mathbf{#1}}
\newcommand{\bs}[1]{\boldsymbol{#1}}

\newcommand{\tiksplotc}[3]{  \raisebox{#1\baselineskip}
	{\scalebox{#2}{ \input{./plots/#3} } }
}

\newcommand{\id}{\mathds{1}}
\newcommand{\vol}{\mathfrak{V}}
%
\title{Supplemental Material: Floquet-Drude conductivity}
\author{Martin Wackerl}
\email{martin.wackerl@ur.de}
\author{Paul Wenk}
\email{paul.wenk@ur.de}
\author{John Schliemann}
\email{john.schliemann@ur.de}
\affiliation{Institute for Theoretical
  Physics, University of Regensburg, 93040 Regensburg, Germany}
%
\date{\today }
%
\maketitle
%
\section{Floquet Kubo formula for the linear conductivity}%
In a driven system the current is not simply a product of resistance
and electric field, see Eq.~(2) together with
Eq.~(3) in the main article. Rather, the conductivity
depends on both the frequency spectrum of the bias, $\omega'$ as well
as a response frequency $\omega$. Furthermore, we introduce
\begin{align}
   \omega  \equiv \tilde{\omega}+p\Omega \quad \Leftrightarrow \quad
  |\tilde\omega| \le \Omega/2\ \qq{with} p\in\mathbb{Z}\;,
\end{align}
and require that the electric field $E^{b}$ depends only
on frequencies $|\tilde{\omega}'|\le\Omega/2$. With this,
Eq.~(2) of the main article becomes
%
\begin{align}
   \langle J^{a}(\f{q},\tilde{\omega}+p\Omega)\rangle = \sum_{b}
   \int_{-\Omega/2}^{\Omega/2}d\tilde{\omega}'\ 
   \bar{\sigma}^{ab}(\f{q},\tilde{\omega}+p\Omega,\tilde{\omega}')E^{b}(\f{q},\tilde{\omega}')
\end{align}
where the conductivity tensor is given by
\begin{align}
   \bar{\sigma}^{ab}(\f{q},\tilde{\omega}+p\Omega,\tilde{\omega}') ={}& \frac{i}{\hbar(\tilde{\omega}+p\Omega)\vol} 
   \sum_{\alpha\beta} \sum_{n_{1}..n_{4}=-\infty}^\infty \frac{f_{\alpha}-f_{\beta}}{\tilde{\omega}+p\Omega+\ih(\ea-\eb) + (n_{1}-n_{2})\Omega+i0^{+}}\nonumber\\
   {}& \times \langle \ua{1} |\, \f{j}^{\ell}(\f{q}) | \ub{2} \rangle\langle \ub{3}  |\, \f{j}^{j}(-\f{q}) | \ua{4} \rangle\nonumber\\
   {}& \times \delta\bigl( \tilde{\omega}+p\Omega + (n_{1}-n_{2}+n_{3}-n_{4})\Omega - \tilde{\omega}' \bigr) \nonumber\\
   {}& +i\frac{e^2 n }{m(\tilde{\omega}+p\Omega)} \delta_{\ell j}\delta(\tilde{\omega}+p\Omega-\tilde{\omega}')\ .
\label{sigmabar}
\end{align}
The argument of the $\delta$-distribution of the first term can become zero if and only if
\begin{align}
   n_{1}-n_{2}+n_{3}-n_{4} = -p\ .
\end{align}
Since we are ultimately interested in the DC limit, we consider only
the case where $p=0$.
Taking only the real part of the longitudinal conductivity one yields
\begin{align}
\begin{aligned}
   \mathrm{Re}\, [\sigma^{xx}(0,\tilde\omega)] = \frac{\pi}{\vol} \left(\frac{e}{m}\right)^{2} \sum_{n,n'}\sum_{\alpha\beta}
   \biggl[ \frac{f_{\alpha}-f_{\beta}}{\hbar\omega} \langle u_{\alpha}^{n} | p^{x} |           u_{\beta}^{n} \rangle
   \langle u_{\beta}^{n'} | p^{x} | u_{\alpha}^{n'} \rangle \,\delta \bigl( \omega +\tfrac{1}{\hbar} (\ea-\eb) \bigr) \biggr]\ .
\end{aligned}
\end{align}
In the limit of $\tilde\omega\to 0$ one ends up with Eq.~(4) of the main article.
\section{Dyson series for time-dependent perturbations}
Here, we use the same notation as in Eqs.~(17) of the
main article. As already shown in the latter, the bare Green's
function $\G_{0}$ fulfills the equation
\begin{align}
   & \Big( i\partial_{t_{1}}-\tfrac{1}{\hbar}
   H_{F}^{0}(\f{x}_{1},t_{1}') \Big)
   \G^{r,a}_{0}(t_{1},t_{2},\f{x}_{1},\f{x}_{2},t_{1}',t_{2}') =\nonumber\\
   & \delta(t_{1}-t_{2})\delta(\f{x}_{1}-\f{x}_{2})\sum_{s=-\infty}^{\infty}\delta(t_{1}'-t_{2}'+sT)  \id
\end{align}
with the definition of the Floquet Hamiltonian for the unperturbed system
\begin{align}
  H_{F}^{0}(\f{x}_{1},t_{1}') = H(\f{x}_{1},t_{1}')-i\hbar\partial_{t_{1}'}
\end{align}
The Hamiltonian for the perturbed one has the form
\begin{align}
   \mathcal{H}_{F}(\f{x}_{1},t_{1},t_{1}') = H_{F}^{0}(\f{x}_{1},t_{1}') + V(\f{x}_{1},t_{1},t_{1}')
\end{align}
where we stress that the dependency on $t_{1}$ is fully kept by the potential. Obviously, the bare Green's function $\G_{0}$ fulfills
\begin{align}
\label{G0}
   &{}\Big( i\partial_{t_{1}}-\tfrac{1}{\hbar}
   \mathcal{H}_{F}(\f{x}_{1},t_{1},t_{1}') +
   \tfrac{1}{\hbar}V(\f{x}_{1},t_{1},t_{1}') \Big)
   \G^{r,a}_{0}(t_{1},t_{2},\f{x}_{1},\f{x}_{2},t_{1}',t_{2}') ={}
   \nonumber\\
   {}& \delta(t_{1}-t_{2})\delta(\f{x}_{1}-\f{x}_{2})\sum_{s=-\infty}^{\infty}\delta(t_{1}'-t_{2}'+sT)  \id\ .
\end{align}
We are interested in the Green's function of the perturbed system $\G_{\mathrm{p}}$ being a solution of 
\begin{align}
\label{Gp}
  {}& \Big( i\partial_{t_{1}}-\tfrac{1}{\hbar}
  \mathcal{H}_{F}(\f{x}_{1},t_{1},t_{1}')\Big)
  \G^{r,a}_{\mathrm{p}}(t_{1},t_{2},\f{x}_{1},\f{x}_{2},t_{1}',t_{2}')
  = \nonumber\\
  {}& \delta(t_{1}-t_{2})\delta(\f{x}_{1}-\f{x}_{2})\sum_{s=-\infty}^{\infty}\delta(t_{1}'-t_{2}'+sT)  \id\ .
\end{align} 
Equating Eq. \eqref{G0} and Eq. \eqref{Gp} we get
\begin{align}
\label{G0=Gp}
   {}& \Big( i\partial_{t_{1}}-\tfrac{1}{\hbar} \mathcal{H}_{F}(\f{x}_{1},t_{1},t_{1}')\Big) \G^{r,a}_{\mathrm{p}}(t_{1},t_{2},\f{x}_{1},\f{x}_{2},t_{1}',t_{2}')  = \nonumber\\
   {}& \Big( i\partial_{t_{1}}-\tfrac{1}{\hbar} \mathcal{H}_{F}(\f{x}_{1},t_{1},t_{1}')\Big) \G^{r,a}_{0}(t_{1},t_{2},\f{x}_{1},\f{x}_{2},t_{1}',t_{2}') +  
   \tfrac{1}{\hbar}V(\f{x}_{1},t_{1},t_{1}')\,\G^{r,a}_{0}(t_{1},t_{2},\f{x}_{1},\f{x}_{2},t_{1}',t_{2}')\ .
\end{align}
The bare Green's function is periodic in both $t_{1}',$ and $t_{2}'$,
\begin{align}
   \G_{0}^{r,a}(t,t_{2},\f{x},\f{x}_{2},t_{1}'+T,t_{2}'+T) = \G_{0}^{r,a}(t,t_{2},\f{x},\f{x}_{2},t_{1}',t_{2}')\ .
\end{align}
Therefore, without loss of generality one can use the restriction
\begin{align}
   t_{1}',t_{2}' \in \big[-\tfrac{T}{2} , \tfrac{T}{2} \big]\ .
\end{align}
Making use of the periodicity of the Green's function and requiring
that the potential is as well periodic in the second time argument
\begin{align}
   V(\f{x},t_{1},t_{1}'+T) = V(\f{x},t_{1},t_{1}')
\end{align}
one can show that
\begin{align}
&   \int_{V_{\f{x}}} d\f{x}\int\limits_{-\infty}^{\infty}dt\int\limits_{-T/2}^{T/2}dt'\ \delta(t_{1}-t)\delta(\f{x}_{1}-\f{x}) \sum_{s=-\infty}^{\infty}\delta(t_{1}'-t'+sT)  V(\f{x},t,t') \G_{0}^{r,a}(t,t_{2},\f{x},\f{x}_{2},t',t_{2}') \nonumber\\
   & = \int\limits_{-T/2}^{T/2}dt'  \sum_{s=-\infty}^{\infty}\delta(t_{1}'-t'+sT)V(\f{x}_{1},t_{1},t') \G_{0}^{r,a}(t_{1},t_{2},\f{x}_{1},\f{x}_{2},t',t_{2}')\\
   & = V(\f{x}_{1},t_{1},t_{1}') \G_{0}^{r,a}(t_{1},t_{2},\f{x}_{1},\f{x}_{2},t_{1}',t_{2}')
\end{align}
where in the last step we have used that the argument of the
delta-distribution can only be zero if $s=0$. Comparing this equation with
Eq.~\eqref{G0=Gp} one finds a Dyson expansion for the Green's function of
the perturbed system
\begin{align}
\begin{aligned} \label{tt'-Dyson}
   \G^{r,a}_{\mathrm{p}}(t_{1},t_{2},\f{x}_{1},\f{x}_{2},t_{1}',t_{2}') =\ & \G^{r,a}_{0}(t_{1},t_{2},\f{x}_{1},\f{x}_{2},t_{1}',t_{2}')\ + \\
   \frac{1}{\hbar}\int_{V_{\f{x}}} d\f{x}\int\limits_{-\infty}^{\infty}dt\int\limits_{-T/2}^{T/2}dt'\ & \G^{r,a}_{\mathrm{p}}(t_{1},t,\f{x}_{1},\f{x},t_{1}',t') V(\f{x},t,t') \G_{0}^{r,a}(t,t_{2},\f{x},\f{x}_{2},t',t_{2}')\ .
\end{aligned}
\end{align}
If one assumes that the potential depends only on the periodic time
component
\begin{align}
   V(\f{x},t,t') = V(\f{x},t') \quad\Leftrightarrow\quad H_{F}(\f{x},t,t') = H_{F}(\f{x},t')
\end{align}
the Green's function depends only on the difference $t_{1}-t_{2}$,
\begin{align}
   \G_{\mathrm{p}}^{r,a}(t_{1},t_{2},\f{x}_{1},\f{x}_{2},t_{1}',t_{2}') =  \G_{\mathrm{p}}^{r,a}(t_{1}-t_{2},\f{x}_{1},\f{x}_{2},t_{1}',t_{2}')\ .
\end{align}
Applying Fourier transform on Eq.~\eqref{tt'-Dyson} with respect to
$t_1 - t_2$, one yields in energy space
\begin{align}
\begin{aligned} \label{Dyson-et'}
   \G^{r,a}_{\mathrm{p}}(\varepsilon,\f{x}_{1},\f{x}_{2},t_{1}',t_{2}') =\ & \G^{r,a}_{0}(\varepsilon,\f{x}_{1},\f{x}_{2},t_{1}',t_{2}')\ + \\
   \ih\int_{V_{\f{x}}} d\f{x}\int\limits_{-T/2}^{T/2}dt'\ & \G^{r,a}_{\mathrm{p}}(\varepsilon,\f{x}_{1},\f{x},t_{1}',t') V(\f{x},t') \G_{0}^{r,a}(\varepsilon,\f{x},\f{x}_{2},t',t_{2}')\ ,
\end{aligned}
\end{align}
where the explicit form of the bare Green's function is given by
\begin{align}
   \G^{r,a}_{0}(\varepsilon,\f{x}_{1},\f{x}_{2},t_{1}',t_{2}') & =    \frac{1}{T}\sum_{r=-\infty}^{\infty}\sum_{\alpha} \frac{u_{\alpha}(\f{x}_{1},t_{1}')\big(u_{\alpha}(\f{x}_{2},t_{2}')\big)^{*}}{\frac{1}{\hbar}\varepsilon-\frac{1}{\hbar}\ea - r\Omega \pm i0}\ e^{ir\Omega(t_{1}'-t_{2}')}\\
   & = \frac{1}{T} \sum_{nn'}\G_{0}^{r,a}(\varepsilon,\f{x}_{1},\f{x}_{2},n,n')e^{-in\Omega t_{1}'}e^{in'\Omega t_{2}'}\;.
\end{align}
The Fourier coefficients are given by
\begin{align}
   \G_{0}^{r,a}(\varepsilon,\f{x}_{1},\f{x}_{2},n,n') = \sum_{r=-\infty}^{\infty}\sum_{\alpha} \frac{u_{\alpha}^{n+r}(\f{x}_{1})\big(u_{\alpha}^{n'+r}(\f{x}_{2})\big)^{*}}{\frac{1}{\hbar}\varepsilon-\frac{1}{\hbar}\ea - r\Omega \pm i0}
\end{align}
where we used the shortened notation
\begin{align}
   u_{\alpha}(\f{x}_{1},t_{1}') \equiv \big\langle \f{x}_{1} | u_{\alpha}(t_{1}') \big\rangle \quad,\quad u_{\alpha}^{n}(\f{x}_{1})\equiv \big\langle \f{x}_{1} | u_{\alpha}^{n} \big\rangle\ .
\end{align}
Since we required the potential to be periodic in the second time
argument it can be expanded in a Fourier series,
\begin{align}
   V(\f{x},t') = \sum_{n=-\infty}^{\infty} V_{n}(\f{x}) e^{-in\Omega t'}\ .
\end{align}
This allows to rewrite Eq.~\eqref{Dyson-et'} and perform the remaining time integration,
\begin{align}
\begin{aligned} \label{Dyson-PeriodicV}
   \G^{r,a}_{\mathrm{p}}(\varepsilon,\f{x}_{1},\f{x}_{2},n,n') =\ & \G^{r,a}_{0}(\varepsilon,\f{x}_{1},\f{x}_{2},n,n')\ + \\
   \ih\int_{V_{\f{x}}} d\f{x} \sum_{n_{1},n_{2}}\ & \G^{r,a}_{\mathrm{p}}(\varepsilon,\f{x}_{1},\f{x},n,n_{1}) V_{n_{1}-n_{2}}(\f{x}) \G_{0}^{r,a}(\varepsilon,\f{x},\f{x}_{2},n_{2},n')\ .
\end{aligned}
\end{align}
\section{$t$-$t^{\prime}$-Formalism}
\subsection{Separating the Periodic from the Aperiodic Time Dependence}
In the $t$-$t^{\prime}$-formalism one starts from the Floquet states
$\ket{\psi_{\alpha}(t)} = \exp(-\frac{i}{\hbar}\varepsilon_{\alpha}t)
  \ket{u_{\alpha}(t)}$ but formally discriminates the time dependence of the
exponential from periodic time dependence as
\begin{equation}
  |\psi_{\alpha}(t^{\prime},t)\rangle=e^{-\frac{i}{\hbar}\varepsilon_{\alpha}t}
  |u_{\alpha}(t^{\prime})\rangle
  \label{ttprime1}
\end{equation}
where obviously
\begin{equation}
  |\psi_{\alpha}(t,t)\rangle=|\psi_{\alpha}(t)\rangle\,.
  \label{ttprime2}
\end{equation}
The advantage of this artifice lies in the fact that the evolution of the
states as a function of $t$ is governed by the operator
\begin{equation}
  U_F(t^{\prime},t)=e^{-\frac{i}{\hbar}H_F(t^{\prime})t}\,,
  \label{ttprime3}
\end{equation}
i.e.
\begin{eqnarray}
  |\psi_{\alpha}(t^{\prime},t)\rangle
  & = &  U_F(t^{\prime},t-t_0)|\psi_{\alpha}(t^{\prime},t_0)\rangle
  \label{ttrprime4}\\
  & = & e^{-\frac{i}{\hbar}\varepsilon_{\alpha}t_0}
  e^{-\frac{i}{\hbar}H_F(t^{\prime})(t-t_0)}|u_{\alpha}(t^{\prime})\rangle
  =e^{-\frac{i}{\hbar}\varepsilon_{\alpha}t}|u_{\alpha}(t^{\prime})\rangle\,,
  \label{ttprime5}
\end{eqnarray}
which avoids any time ordering.

On the space of all states depending periodically with period $T$ on a
parameter $t^{\prime}$ having dimension of time, we define the saclar product
\begin{equation}
  (\varphi|\chi)
  = \frac{1}{T} \int_0^Tdt^{\prime}\,\langle\varphi(t^{\prime})|\chi(t^{\prime})\rangle
  = \frac{1}{T} \int_0^Tdt^{\prime}\,\langle\varphi|t^{\prime}\rangle\langle t^{\prime}|\chi\rangle
  \label{ttprime6}
\end{equation}
which differs from the scalar product introduced by Sambe\cite{Sambe73}
\begin{align}
  \langle\!\langle\varphi(t)|\chi(t)\rangle\!\rangle
  =\frac{1}{T}\int_0^Tdt\,\langle\varphi(t)|\chi(t)\rangle
\end{align}
by a factor $1/T$. The notation
\begin{equation}
  \langle t^{\prime}|\psi\rangle:=|\psi(t^{\prime})\rangle
  \label{ttprime7}
\end{equation}
suggests to consider $t^{\prime}$ as a coordinate rather than a time parameter.
The corresponding operator $\hat t^{\prime}$ can be defined to act
multiplicatively on the above wave functions,
\begin{equation}
  \hat t^{\prime}|\psi(t^{\prime})\rangle
  =\langle t^{\prime}|\hat t^{\prime}|\psi\rangle
  =t^{\prime}|\psi(t^{\prime})\rangle\,,
  \label{ttprime8}
\end{equation}
and the canonically conjugate operator is
\begin{equation}
  \hat w:=-i\hbar\partial_{t^{\prime}}=H_F(t^{\prime})-h(t^{\prime})
  \quad\Rightarrow\quad\left[\hat w,\hat t^{\prime}\right]=-i\hbar
  \label{ttprime9}
\end{equation}
with a complete system of orthonormalized periodic eigenfunctions
\begin{equation}
  \langle t^{\prime}|l\rangle = e^{-i\Omega lt^{\prime}}
  \quad,\quad\hat w|l\rangle=l\hbar\Omega|l\rangle
  \quad,\quad(k|l)=\delta_{kl},\qq{with}k,l\in\mathbb{Z}\,,
  \label{ttprime10}
\end{equation}
\begin{equation}
  \sum_{l=-\infty}^{\infty}\langle t^{\prime}_1|l\rangle\langle l|t^{\prime}_2\rangle
  = T\sum_{s=-\infty}^{\infty}\delta(t^{\prime}_1-t^{\prime}_2+sT)
  = \langle t^{\prime}_1|t^{\prime}_2\rangle\,.
  \label{ttprime11}
\end{equation}
In obtaining the completeness relation we have taken
into account the Fourier expansion of the Dirac comb,
\begin{equation}
  \sum_{r=-\infty}^{\infty}e^{ir\Omega t}
  = T \sum_{s=-\infty}^{\infty}\delta(t+sT)\,.
  \label{ttprime12}
\end{equation}
Switching between the two pertaining representations
amounts, up to signs and prefactors, in the usual Fourier expansion,
\begin{align}
  \langle l|\psi\rangle ={}&  \frac{1}{T}\int_0^Tdt^{\prime}\,\langle l|t^{\prime}\rangle
  \langle t^{\prime}|\psi\rangle
  =\frac{1}{T}
  \int_0^Tdt^{\prime}\,e^{i\Omega lt^{\prime}}\langle t^{\prime}|\psi\rangle
  \label{ttprime13}\\
  \qq{$\Leftrightarrow$}
  \langle t^{\prime}|\psi\rangle
  ={}& \sum_{l=-\infty}^{\infty}\langle t^{\prime}|l\rangle
  \langle l|\psi\rangle
  =\sum_{l=-\infty}^{\infty}
  e^{-i\Omega lt^{\prime}}\langle l|\psi\rangle\,.
  \label{ttprime14}
\end{align}
Finally, the analogs of the wave functions
$\psi_{\alpha}(q,t) = \braket{q}{\psi_{\alpha}(t)}$ read in the
$t$-$t^{\prime}$-formalism
\begin{equation}
  \psi_{\alpha}(q,t^{\prime},t)
  =\langle q|\psi_{\alpha}(t^{\prime},t)\rangle
  =\langle q,t^{\prime}|\psi_{\alpha}(t)\rangle\,.
  \label{ttprime15}
\end{equation}
\subsection{Field Operators and One-Particle Green's Functions}
Generalizing the states (\ref{ttprime1}) we define
\begin{equation}
  |\phi^r_{\alpha}(t^{\prime},t)\rangle
  = e^{ir\Omega(t^{\prime}-t)}|\psi_{\alpha}(t^{\prime},t)\rangle
  = e^{-\frac{i}{\hbar}(\varepsilon_{\alpha}+r\hbar\Omega)t}
  e^{ir\Omega t^{\prime}}|u_{\alpha}(t^{\prime})\rangle
  \label{ttprime16}
\end{equation}
with
\begin{equation}
  \phi^r_{\alpha}(q,t^{\prime},t)=\langle q|\phi^r_{\alpha}(t^{\prime},t)\rangle
  =\langle q,t^{\prime}|\phi^r_{\alpha}(t)\rangle
  \label{ttprime17}
\end{equation}
and the simple properties
\begin{equation}
  |\phi^r_{\alpha}(t^{\prime},t)\rangle
  =U_F(t^{\prime},t-t_0)|\phi^r_{\alpha}(t^{\prime},t_0)\rangle\,,
  \label{ttprime18}
\end{equation}
\begin{equation}
  (\phi^r_{\alpha}(t)|\phi^s_{\beta}(t))=\delta_{\alpha\beta}\delta_{rs}\,,
  \label{ttprime19}
\end{equation}
\begin{equation}
  \sum_{\alpha}\sum_{r=-\infty}^{\infty}
  |\phi^r_{\alpha}(t^{\prime}_1,t)\rangle\langle\phi^r_{\alpha}(t^{\prime}_2,t)|
  ={} \id\, T\sum_{s=-\infty}^{\infty}\delta(t^{\prime}_1-t^{\prime}_2+sT)\,.
  \label{ttprime20}
\end{equation}
In second quantization this allows to define a system of 
creation and annihilation operators $b^\dagger_{\alpha r}(t)$, $b_{\alpha r}(t)$ with
\begin{equation}
  |\phi^r_{\alpha}(t)\rangle=b^\dagger_{\alpha r}(t)|0\rangle\qquad,\qquad
  b_{\alpha r}(t)|0\rangle=0
  \label{ttprime21}
\end{equation}
and
\begin{equation}
  \left[b_{\alpha r}(t),b^\dagger_{\beta s}(t)\right]_{\pm}=\delta_{\alpha\beta}\delta_{rs}
  \quad,\quad
  \left[b_{\alpha r}(t),b_{\beta s}(t)\right]_{\pm}
  =\left[b^\dagger_{\alpha r}(t),b^\dagger_{\beta s}(t)\right]_{\pm}=0\,.
  \label{ttprime22}
\end{equation}
Field operators can be constructed as
\begin{equation}
  \Phi(q,t^{\prime},t)=\sum_{\alpha}\sum_{r=-\infty}^{\infty}
  \phi_{\alpha}(q,t^{\prime},t)b_{\alpha r}(t)
  \label{ttprime23}
\end{equation}
fulfilling
\begin{equation}
  \left[\Phi(q_1,t^{\prime}_1,t),\Phi^\dagger(q_1,t^{\prime}_2,t)\right]_{\pm}
  =\delta(q_1-q_2)\, T\sum_{s=-\infty}^{\infty}\delta(t^{\prime}_1-t^{\prime}_2+sT)
  \label{ttprime24}
\end{equation}
with again all other (anti-)commutators at equal times $t$ being zero, and
\begin{equation}
  \langle q_2,t^{\prime}_2|\Phi^\dagger(q_1,t^{\prime}_1,t)|0\rangle
  =\delta(q_1-q_2)\, T\sum_{s=-\infty}^{\infty}\delta(t^{\prime}_1-t^{\prime}_2+sT)\,.
  \label{ttprime25}
\end{equation}
The Floquet Hamiltonian $H_F$ can be formulated as
\begin{align}
  H_F(t) ={}& \sum_{\alpha}\sum_{r=-\infty}^{\infty}
  \left(\varepsilon_{\alpha}+r\hbar\Omega\right)b^\dagger_{\alpha r}(t)b_{\alpha r}(t)
  \label{ttprime26}
\end{align}
and is neither bounded form below nor from above. Going over to the
Heisenberg picture,
\begin{equation}
  \Phi_H(q,t^{\prime},t)
  =U^\dagger_F(t^{\prime},t)\Phi(q,t^{\prime},t)U_F(t^{\prime},t)
  =\sum_{\alpha}\sum_{r=-\infty}^{\infty}
  \phi_{\alpha}(q,t^{\prime},t)b_{\alpha r}(0)\,,
  \label{ttprime27}
\end{equation}
we yield the retarded/advanced one-particle Green's function
\begin{eqnarray}
  & & {\cal G}^{r/a}(q_1,t^{\prime}_1,t_1,q_2,t^{\prime}_2,t_2)\nonumber\\
  & & \qquad\qquad =\mp i\Theta(\pm(t_1-t_2))\frac{1}{T}
  \left\langle\left[\Phi_H(q_1,t^{\prime}_1,t_1)
    ,\Phi^\dagger_H(q_2,t^{\prime}_2,t_2)\right]_{\epsilon}\right\rangle
  \label{ttprime28}\\
  & & \qquad\qquad =\mp i\Theta(\pm(t_1-t_2))\frac{1}{T}
  \sum_{r=-\infty}^{\infty}\sum_{\alpha}\phi^r_{\alpha}(q_1,t^{\prime}_1,t_1)
  \left(\phi^r_{\alpha}(q_2,t^{\prime}_2,t_2)\right)^{\ast}
  \label{ttprime29}\\
  & & \qquad\qquad
  =\mp i\Theta(\pm(t_1-t_2))\frac{1}{T}\sum_{r=-\infty}^{\infty}\sum_{\alpha}
  \Bigl[e^{-\frac{i}{\hbar}(\varepsilon_{\alpha}+r\hbar\Omega)(t_1-t_2)}\nonumber\\
    & & \qquad\qquad\qquad\qquad
    \cdot\langle q_1|u_{\alpha}(t^{\prime}_1)\rangle
    \langle u_{\alpha}(t^{\prime}_2)|q_2\rangle
    e^{ir\Omega(t^{\prime}_1-t^{\prime}_2)}\Bigr]\,,
  \label{ttprime30}
\end{eqnarray}
or, formulated as a Green's operator,
\begin{eqnarray}
  & & \hat{\cal G}^{r/a}(t^{\prime}_1,t_1,t^{\prime}_2,t_2)\nonumber\\
   & & \qquad\qquad =\mp i\Theta(\pm(t_1-t_2))\frac{1}{T}
  \sum_{r=-\infty}^{\infty}\sum_{\alpha}|\phi^r_{\alpha}(t^{\prime}_1,t_1)\rangle
  \langle\phi^r_{\alpha}(t^{\prime}_2,t_2)|
  \label{ttprime31}\\
   & & \qquad\qquad
  =\mp i\Theta(\pm(t_1-t_2))\frac{1}{T}\sum_{r=-\infty}^{\infty}\sum_{\alpha}
  \Bigl[e^{-\frac{i}{\hbar}(\varepsilon_{\alpha}+r\hbar\Omega)(t_1-t_2)}\nonumber\\
    & & \qquad\qquad\qquad\qquad
    \cdot|u_{\alpha}(t^{\prime}_1)\rangle
    \langle u_{\alpha}(t^{\prime}_2)|
    e^{ir\Omega(t^{\prime}_1-t^{\prime}_2)}\Bigr]\,.
  \label{ttprime32}
\end{eqnarray}
These quantities have the significant property
\begin{align}
  \left(i\partial_{t_1}-\frac{1}{\hbar}H_F(t^{\prime}_1)\right)
  \hat{\cal G}^{r/a}(t_1,t_2,t^{\prime}_1,t^{\prime}_2)
   ={}& \delta(t_1-t_2)\sum_{\alpha}
  |u_{\alpha}(t^{\prime}_1)\rangle\langle u_{\alpha}(t^{\prime}_2)|
  \frac{1}{T}\sum_{r=-\infty}^{\infty}e^{ir\Omega(t^{\prime}_1-t^{\prime}_2)}\nonumber\\
  ={}& \delta(t_1-t_2)\id
  \sum_{s=-\infty}^{\infty}\delta(t^{\prime}_1-t^{\prime}_2+sT)
  \label{ttprime33}
\end{align}
where we have used Eqs.~(\ref{ttprime12}), and the completeness
relation of the Floquet functions $\ket{u_\alpha(t)}$. As the expressions
(\ref{ttprime31}), (\ref{ttprime32}) depend only on the difference
$t_1-t_2$ and are periodic in $t^{\prime}_1$, $t^{\prime}_2$ we can go over
to Fourier components as
\begin{eqnarray}
  \hat{\cal G}^{r/a}(\omega,n_1,n_2) & = & \int_{-\infty}^{\infty}dt\,
  e^{i\omega t}\sum_{n_1,n_2=-\infty}^{\infty}
  e^{-i n_1\Omega t^{\prime}_1 + i n_2\Omega t^{\prime}_2}
  \hat{\cal G}^{r/a}(t^{\prime}_1,t,t^{\prime}_2,0)
  \label{ttprime34}\\
  & = & \frac{1}{T}\sum_{r=-\infty}^{\infty}\sum_{\alpha}
  \frac{|u^{n_1 + r}_{\alpha}\rangle\langle u^{n_2 + r}_{\alpha}|}
       {\omega-\frac{1}{\hbar}(\varepsilon_{\alpha}+r\hbar\Omega)\pm i
        0^+}
        \label{ttprime35}
\end{eqnarray}
where the last line follows from
\begin{equation}
  \Theta(\pm t)=\frac{\pm i}{2\pi}\int_{-\infty}^{\infty}d\omega\,
  \frac{e^{-i\omega t}}{\omega\pm i 0^+}\,.
  \label{ttprime36}
\end{equation}
Moreover, with the spectral density
\begin{eqnarray}
  A(q_1,t^{\prime}_1,t_1,q_2,t^{\prime}_2,t_2)
  & = & \frac{1}{2\pi T}\left\langle\left[\Phi_H(q_1,t^{\prime}_1,t_1)
    ,\Phi^\dagger_H(q_2,t^{\prime}_2,t_2)\right]_{\epsilon}\right\rangle
  \label{ttprime37}\\
  & = & \frac{1}{2\pi T}\sum_{r=-\infty}^{\infty}\sum_{\alpha}
  \Bigl[e^{-\frac{i}{\hbar}(\varepsilon_{\alpha}+r\hbar\Omega)(t_1-t_2)}\nonumber\\
    & & \qquad\qquad
    \cdot\langle q_1|u_{\alpha}(t^{\prime}_1)\rangle
    \langle u_{\alpha}(t^{\prime}_2)|q_2\rangle
    e^{ir\Omega(t^{\prime}_1-t^{\prime}_2)}\Bigr]\,,
  \label{ttprime38}
\end{eqnarray}
having Fourier components
\begin{eqnarray}
  A(\omega,q_1,t^{\prime}_1,q_2,t^{\prime}_2)
  & = & \frac{1}{T}\sum_{r=-\infty}^{\infty}\sum_{\alpha}
  \Bigl[\delta\left(\omega-(\varepsilon_{\alpha}+r\hbar\Omega)\right)\nonumber\\
    & & \qquad\qquad
    \cdot\langle q_1|u_{\alpha}(t^{\prime}_1)\rangle
    \langle u_{\alpha}(t^{\prime}_2)|q_2\rangle
    e^{ir\Omega(t^{\prime}_1-t^{\prime}_2)}\Bigr]\,,
  \label{ttprime39}
\end{eqnarray}
we obtain the familiar Lehmann representation of the Green's function,
\begin{equation}
  {\cal G}^{r/a}(\omega,q_1,t^{\prime}_1,q_2,t^{\prime}_2)
  =\int_{-\infty}^{\infty}d\omega^{\prime}\,
  \frac{A(\omega,q_1,t^{\prime}_1,q_2,t^{\prime}_2)}
       {\omega-\omega^{\prime}\pm i 0^+}\,.
       \label{ttprime40}
\end{equation}
In summary, when treating $t^{\prime}$ not as a time parameter but rather as a
state coordinate, the remaining time evolution in $t$ is governed by the
Floquet Hamiltonian being independent of $t$. Thus, we are left with an
effectively time-independent Hamiltonian, and many formal manipulations known
for such a situation work just in the same way. Note, however, that
(i) the physical case is still requires $t=t^{\prime}$, and (ii) the Floquet
Hamiltonian (\ref{ttprime26}) fails to be bounded from below.
\section{Floquet Born approximation and generalized Floquet Fermi's golden rule}
The aim of this section is to relate the self-energy in first order
Born approximation (1BA) to the scattering time given by Fermi's
golden rule~\cite{akkermans,Dirac27,Fermi50,Mahan2013,sakurai}. This requires Fermi's golden rule to be applicable to $tt'$-Floquet
states. To do so, let us first a recall Fermi's golden rule for Floquet states.
\subsection{Floquet Fermi's Golden Rule}
A generalization of Fermi's golden rule to time periodic
Hamiltonians, i.e. the Floquet Fermi golden rule, was already derived
by Kitagawa \textit{et. al.} in Ref.~\cite{Kitagawa11}. However, a
detailed derivation and discussion of the \grqq Scattering theory for
Floquet-Bloch states\grqq\,, is given in Ref.~\cite{Bilitewski15}. The
Floquet Fermi's golden rule was used by O.V. Kibis in
Ref.~\cite{Kibis14} in order to explain the suppression of
backscattering of conduction electrons in presence of a high-frequency
electric field.  In regards to Fermi's golden rule for the
$tt'$-Floquet states, the derivation of the Floquet Fermi's golden rule
is presented here in detail. It is assumed that the solution of the
time-dependent Schr\"odinger equation
\begin{align}\label{FermiSchrödinger}
i\hbar\frac{\partial}{\partial t} |\psi_{\alpha}(t) \rangle = H(t) |\psi_{\alpha}(t) \rangle
\end{align}
and the corresponding time evolution operator $U_{0}(t,t_{0})$ are
known. In presence of a time-dependent perturbation $V(t)$ the
Schr\"odinger equation becomes
\begin{align}\label{FermiSchrödingerV}
i\hbar\frac{\partial}{\partial t} |\Psi_{\alpha}(t) \rangle = \big[H(t) + V(t) \big] |\Psi_{\alpha}(t) \rangle\ .
\end{align}
The potential $V(t)$ is switched on at a reference time $t_{0}$ such that the solutions of the Schr\"odinger equation coincide for times $t\le t_{0}$
\begin{align}\label{FermipsiPsi}
|\psi_{\alpha}(t) \rangle = |\Psi_{\alpha}(t) \rangle \quad \text{for}\quad t\le t_{0}\ .
\end{align}
At times $t\le t_{0}$ the particle is assumed to be in an eigenstate
of the unperturbed Hamiltonian.
Standard perturbation theory leads to the transition amplitude
\begin{align}\label{FermiFirstOrder}
\langle \psi_{\beta}(t) | \Psi_{\alpha}(t) \rangle =  \delta_{\alpha\beta} + \frac{1}{i\hbar} \int\limits_{t_{0}}^{t}dt'\, \langle \psi_{\beta}(t') | V(t') | \psi_{\alpha}(t') \rangle
\end{align}
up to first order in the potential. Without loss of generality $t_{0}$
can be set to zero and for $\alpha\ne\beta$ the first nontrivial order
of Eq.~\eqref{FermiFirstOrder} is given by
\begin{align}\label{Fermia(t)}
a_{\alpha\beta}(t) = -\frac{i}{\hbar} \int\limits_{0}^{t}dt'\,\langle \psi_{\beta}(t') | V(t') | \psi_{\alpha}(t') \rangle\ .
\end{align}
This formula, the Floquet Fermi's golden rule, is equal to Eq.~(10) of
Ref.~\cite{Kibis14}. To proceed further, scattering from a Floquet
state into a state with constant energy
\begin{align}
|\psi_{\alpha}(\varepsilon,t) \rangle = e^{-\frac{i}{\hbar}\varepsilon t}|u_{\alpha}(t) \rangle
\end{align}
is considered. The quasienergy $\varepsilon$ is independent of the quantum number. Hence, this state is not an eigenstate of the Hamiltonian, nevertheless it fulfills
\begin{align}
\langle \psi_{\alpha}(t) | \psi_{\beta}(\varepsilon,t) \rangle = \delta_{\alpha\beta}\,e^{\frac{i}{\hbar}(\ea-\varepsilon)t}\ .
\end{align}
Consequently, Eq.~\eqref{Fermia(t)} remains valid if the final state
is $|\psi_{\alpha}(\varepsilon,t) \rangle$. Now, consider a scattering
event from a Floquet state $\psi_{\alpha}(\f{k}',t)$ into a state with
constant energy $\varepsilon$,
\begin{align}\label{ScatteringFloquetConstantEnergy}
\psi_{\alpha}(\f{k}',t) = e^{-\frac{i}{\hbar}\ea(\f{k}') t}u_{\alpha}(\f{k}',t) \leadsto e^{-\frac{i}{\hbar}\varepsilon t} u_{\beta}(\f{k},t)\ .
\end{align}
If the perturbation $V(t)$ is time-independent Eq.~\eqref{Fermia(t)} becomes
\begin{align}
a_{\alpha\beta}(\f{k},\f{k}',t) = -i\frac{V_{\f{k}\f{k}'}}{\hbar}\sum_{nn'=-\infty}^{\infty}\int\limits_{0}^{t}dt'\, e^{\frac{i}{\hbar}(\varepsilon-\ea(\f{k}')-(n-n')\hbar\Omega)t'} \big(u_{\beta}^{n'}(\f{k})\big)^{*} u_{\alpha}^{n}(\f{k}')
\end{align}
with
$V_{\f{k}\f{k}'} =
\mel{\varphi_{\vb{k},\vb{r}}}{V(\vb{r})}{\varphi_{\vb{k}',\vb{r}}}$
where
$\varphi_{\vb{k},\vb{r}} = \exp(-i \vb{k}\cdot
\vb{r})/\sqrt{\vol}$. Shifting $t'$ by $-t/2$ yields the probability
density
\begin{align}\label{FermiIntermediateStep}
\begin{aligned}
|a_{\alpha\beta}(\f{k},\f{k}',t)|^{2} = \frac{V_{\f{k}\f{k}'}^{2}}{\hbar^{2}} \bigg| & \sum_{nn'=-\infty}^{\infty} e^{\frac{i}{2\hbar}(\varepsilon-\ea(\f{k}')-(n-n')\hbar\Omega)t} \big(u_{\beta}^{n'}(\f{k})\big)^{*} u_{\alpha}^{n}(\f{k}') \\ 
& \times\,\int\limits_{-t/2}^{t/2}dt'\, e^{\frac{i}{\hbar}(\varepsilon-\ea(\f{k}')-(n-n')\hbar\Omega)t'} \bigg|^{2}\ .
\end{aligned}
\end{align}
In the long time limit $t \rightarrow \infty$ this simplifies to
%
\begin{align}\label{aalpha}
\begin{aligned}
|a_{\alpha\beta}(\f{k},\f{k}',t)|^{2} = 4\pi^{2}V_{\f{k}\f{k}'}^{2} \bigg| \sum_{nn'=-\infty}^{\infty} &   \big(u_{\beta}^{n'}(\f{k})\big)^{*} u_{\alpha}^{n}(\f{k}')
\delta\big(\varepsilon-\ea(\f{k}')-(n-n')\hbar\Omega\big)\bigg|^{2}\ .
\end{aligned}
\end{align}
The quasienergies $\varepsilon$ and $\ea(\vb{k})$ are chosen to be in the
central Floquet zone such that
\begin{align}
\forall_{\vb{k}}:\;|\varepsilon-\ea(\vb{k} )| \le \hbar\Omega\, ,\\
\begin{split}
\delta(\varepsilon-\ea-n\hbar\Omega)\delta(\varepsilon-\ea-m\hbar\Omega) &= \delta^{2}(\varepsilon-\ea-n\hbar\Omega)\delta_{nm}\ . \label{FermiProductDelta}
\end{split}
\end{align}
Hence, Eq.~\eqref{aalpha} becomes
\begin{align}
|a_{\alpha\beta}(\f{k},\f{k}',t)|^{2} = 4\pi^{2}V_{\f{k}\f{k}'}^{2} \sum_{n=-\infty}^{\infty}  c_{\beta\alpha}^{-n}(\f{k},\f{k}') \big(c_{\beta\alpha}^{-n}(\f{k},\f{k}')\big)^{*}\, \delta^{2}\big(\varepsilon-\ea(\f{k}')-n\hbar\Omega\big)\ .
\end{align}
with
\begin{align}\label{ccoefficient}
   c_{\alpha\beta}^{n}(\f{k},\f{k'}) \equiv \sum_{m=-\infty}^{\infty}  \big(u_{\alpha}^{m+n}(\f{k})\big)^{*} u_{\beta}^{m}(\f{k}') \ .
\end{align}

The square of the delta-distribution can be rewritten as \cite{Kibis14}
\begin{align}\label{FermiDeltaSquared}
\delta^{2}(\varepsilon) = \delta(\varepsilon)\delta(0)=\frac{\delta(\varepsilon)}{2\pi\hbar}\,\lim_{t\to\infty}\int\limits_{-t/2}^{t/2}dt'\,e^{\frac{i}{\hbar}0t'} = \frac{\delta(\varepsilon)t}{2\pi\hbar}\ .
\end{align}
The transition probability is then
\begin{align}
\Gamma_{\alpha\beta}(\f{k},\f{k}') & \equiv \frac{d|a_{\alpha\beta}(\f{k},\f{k}',t)|^{2}}{dt} \\
& = \frac{2\pi}{\hbar} V_{\f{k}\f{k}'}^{2}\sum_{n=-\infty}^{\infty}  c_{\beta\alpha}^{-n}(\f{k},\f{k}') \big(c_{\beta\alpha}^{-n}(\f{k},\f{k}')\big)^{*}\,\delta\big(\varepsilon-\ea(\f{k}')-n\hbar\Omega\big)\ .
\end{align}
The delta-distribution can only have support if $n=0$. Performing an impurity average according to the main article, leads to  $\langle V_{\f{k}\f{k}'}^{2}\rangle_{\mathrm{imp}}=\Vimp$ such that
\begin{align}
\langle\Gamma_{\alpha\beta}(\f{k},\f{k}')\rangle_{\mathrm{imp}} & = \langle\Gamma_{\alpha\beta}(\f{k},\f{k}')\rangle_{\mathrm{imp}} \\
& = \frac{2\pi}{\hbar} \Vimp |c_{\beta\alpha}^{0}(\f{k},\f{k}')|^{2}\delta\big(\varepsilon-\ea(\f{k}')\big)\ .
\end{align}
The scattering time is then governed by the sum over all initial states and the sum over all momenta
\begin{align}\label{ScatteringTimeGoldenRule}
\frac{1}{\tau_{\beta}(\varepsilon,\f{k})} & = \frac{1}{V_{\f{k}'}}\sum_{\f{k}'}\sum_{\alpha} \langle\Gamma_{\alpha\beta}(\f{k},\f{k}')\rangle_{\mathrm{imp}} \\
& =  \frac{2\pi}{\hbar}\Vimp \frac{1}{V_{\f{k}'}}\sum_{\f{k}'}\sum_{\alpha} |c_{\beta\alpha}^{0}(\f{k},\f{k}')|^{2}\,\delta\big(\varepsilon-\ea(\f{k}') \big)\;.
\end{align}
The last equation is the Floquet Fermi's golden rule.
\subsection{Fermi's Golden Rule for $tt'$-Floquet States}
In the following the steps of the derivation of the Fermi's Golden
Rule for $tt'$-Floquet states are similar to the one
applied in Refs.~\cite{Mahan2013,Kibis14}. The difference lies in the use of the
$tt'$-Floquet states (see Eq.~(7) in the main article)
instead of the Floquet states. A $tt'$-state fulfills
\begin{align}
i\hbar\frac{\partial}{\partial t} |\psi^{\ell}_{\alpha}(t,t') \rangle = H_{F}(t') |\psi^{\ell}_{\alpha}(t,t') \rangle\ .
\end{align}
The corresponding time-evolution operator fulfilling this
Schr\"odinger equation is given by
\begin{align}
U_{0}(t,t_{0},t') = e^{-\frac{i}{\hbar}H_{F}(t')\cdot(t-t_{0})}\ .
\end{align}
If a perturbation is switched on at time $t_{0}$ the Schr\"odinger equation becomes
\begin{align}\label{tt'Schrodinger}
i\hbar\frac{\partial}{\partial t} |\Psi^{\ell}_{\alpha}(t,t') \rangle = \big[ H_{F}(t') + V(t,t') \big] |\Psi^{\ell}_{\alpha}(t,t') \rangle
\end{align}
with the boundary condition $|\psi^{\ell}_{\alpha}(t,t')\rangle =
|\Psi^{\ell}_{\alpha}(t,t')\rangle \quad \text{for} \quad t\le t_{0}\ $.
Changing into the interaction picture with
\begin{align}\label{tt'InteractionState}
|\Psi^{\ell}_{\alpha}(t,t')\rangle_{I} ={}&  U_{0}^{\dagger}(t,t_{0},t')
  |\Psi^{\ell}_{\alpha}(t,t')\rangle\:,\\
  V_{I}(t,t') ={}&  U_{0}^{\dagger}(t,t_{0},t') V(t,t') U_{0}(t,t_{0},t')\ ,
\end{align}
%
%
one finds up to first order in the potential $V$
\begin{align}\label{tt'FermiFirstOrder}
|\Psi^{\ell}_{\alpha}(t,t') \rangle_{I}  \approx{}&
  |\psi^{\ell}_{\alpha}(t_{0},t') \rangle + \frac{1}{i\hbar}
  \int_{t_{0}}^{t}dt_{1}\,
  V_{I}(t_{1},t')|\psi^{\ell}_{\alpha}(t_{0},t') \rangle\;\\
  \langle \psi^{\ell'}_{\beta}(t,t'') | \Psi_{\alpha}^{\ell}(t,t') \rangle ={}& \, \langle \psi^{\ell'}_{\beta}(t,t'') | \psi_{\alpha}^{\ell}(t,t') \rangle \\
&+ \frac{1}{i\hbar} \int_{t_{0}}^{t}dt_{1}\, \langle \psi^{\ell'}_{\beta}(t_{1},t'') | V(t_{1},t')|\psi_{\alpha}^{\ell}(t_{1},t') \rangle\ .
\end{align}
In the next step let us consider the matrix element where the $tt'$-Floquet states have the same time dependence but different Floquet indices,
\begin{align}
a^{\ell\ell'}_{\alpha\beta}(t,t') & = \sum_{n=-\infty}^{\infty} a^{\ell\ell'}_{\alpha\beta}(t,n)\,e^{in\Omega t'}\\
& = \langle \psi_{\beta}^{\ell}(t,t') | \Psi_{\alpha}^{\ell'}(t,t') \rangle\\
\begin{split}
& \approx \delta_{\alpha\beta}\, e^{i\Omega(\ell-\ell')(t'-t)}+ \frac{1}{i\hbar}\int\limits_{t_{0}}^{t}dt_{1}\, \langle \psi_{\beta}^{\ell}(t_{1},t') |V(t_{1},t')|\psi_{\alpha}^{\ell'}(t_{1},t')\rangle\;. \label{Fermitt'leadingorder}                
\end{split}
\end{align}
The Fourier coefficients for a perturbation, which is time-independent
in the second time argument, are governed by
\begin{align}
a^{\ell\ell'}_{\alpha\beta}(t,n) & = \frac{1}{T}\int\limits_{0}^{\frac{2\pi}{T}}dt'\, a^{\ell\ell'}_{\alpha\beta}(t,t')\,e^{in\Omega t'} \\
\begin{split}
& = \delta_{\alpha\beta}\delta_{n,\ell-\ell'}\,e^{-in\Omega t}\, + \frac{1}{i\hbar}\int_{t_{0}}^{t}dt_{1}\, e^{\frac{i}{\hbar}\big(\ea-\eb+(\ell-\ell')\hbar\Omega\big)t_{1}} \sum_{m=-\infty}^{\infty} \langle u_{\alpha}^{m+\ell+n} |V(t_{1})| u_{\beta}^{m+\ell'}\rangle\;.
\end{split}
\end{align}
We see that the transition amplitude is
only a function of the difference of the Floquet indices, $a_{\alpha\beta}^{\ell\ell'}(t,t') = a_{\alpha\beta}^{(\ell-\ell')}(t,t')$.
Analogue to the last section, $t_{0}$ can be set to zero and for
$\alpha\ne\beta$  Eq.~\eqref{Fermitt'leadingorder} simplifies to
\begin{align}\label{AfirstOrder}
a_{\alpha\beta}^{\ell\ell'}(t,t') = -\frac{i}{\hbar} \int\limits_{0}^{t}dt_{1}\, \langle \psi_{\beta}^{\ell}(t_{1},t') |V(t_{1},t')|\psi_{\alpha}^{\ell'}(t_{1},t')\rangle\ .
\end{align}
Now, let us assume a scattering event from a $tt'$-Floquet state into
another $tt'$-Floquet state with constant quasienergy, given by
\begin{align}\label{StateWithConstEnergy}
|\psi_{\alpha}^{\ell}(\varepsilon,t,t')\rangle \equiv e^{-\frac{i}{\hbar}(\varepsilon+\ell\hbar\Omega) t} |u_{\alpha}(t,t')\rangle e^{i\ell\Omega t'}\;.
\end{align}
The quasienergy is independent of the quantum number. This state is not an eigenstate of the Hamiltonian, nevertheless it fulfills
\begin{align}
\langle \psi_{\alpha}^{\ell}(t,t') | \psi_{\beta}^{\ell'}(\varepsilon,t,t') \rangle = \delta_{\alpha\beta} e^{-\frac{i}{\hbar}(\varepsilon-\ea+(\ell'-\ell)\hbar\Omega)t}\,e^{i\Omega(\ell'-\ell)t'}\ .
\end{align}
Hence, Eq.~\eqref{AfirstOrder} remains valid if the final state is of
the same form as in Eq.~\eqref{StateWithConstEnergy}. Consider now a
scattering event from a $tt'$-Floquet state $\psi_{\alpha}^{\ell}(\f{k}',t,t')$ into a state with constant
energy $\varepsilon$,
\begin{align}\label{tt'stateconstenergy}
\psi_{\alpha}^{\ell}(\f{k}',t,t') = e^{-\frac{i}{\hbar}(\ea(\f{k}') + \ell\hbar\Omega ) t} u_{\alpha}(\f{k}',t')\,^{i\ell\Omega t'} \leadsto e^{-\frac{i}{\hbar}(\varepsilon + \ell'\hbar\Omega ) t} u_{\beta}(\f{k},t')\,^{i\ell'\Omega t'}\ .
\end{align}
The Fourier coefficient of the matrix element for a scattering as in
Eq.~\eqref{tt'stateconstenergy} is for a time-independent perturbation
given by
\begin{align}
\begin{split}
a_{\alpha\beta}^{\ell\ell'}(\f{k},\f{k}',t,n) = & -i\frac{V_{\f{kk}'}}{\hbar}\int\limits_{0}^{t}dt'\, e^{\frac{i}{\hbar}(\varepsilon-\ea(\f{k}')-(\ell-\ell')\hbar\Omega)t'}\\
&\hspace{0.5cm}\times\, \sum_{m=-\infty}^{\infty} \big(u_{\beta}^{m+\ell+n}(\f{k})\big)^{*}u_{\alpha}^{m+\ell'}(\f{k}')
\end{split}\\
= & -i\frac{V_{\f{kk}'}}{\hbar}\int\limits_{0}^{t}dt'\, e^{\frac{i}{\hbar}(\varepsilon-\ea(\f{k}')-(\ell-\ell')\hbar\Omega)t'} c_{\beta\alpha}^{\ell-\ell'+n}(\f{k},\f{k}')\;.
\end{align}
In the last step the definition given in Eq.~\eqref{ccoefficient} has
been used. This allows for a definition of the transition probability matrix
\begin{align}
\big(\f{A}_{\alpha\beta}^{\ell\ell'jj'}(\f{k},\f{k}',t)\big)_{n,n'} \equiv \sum_{\gamma} a_{\gamma\alpha}^{\ell\ell'}(\f{k},\f{k}',t,n)\big( a_{\gamma\beta}^{jj'}(\f{k},\f{k}',t,n') \big)^{*}\ .
\end{align}
Equivalently to Eq.~\eqref{FermiIntermediateStep}, in the limit $t
\rightarrow \infty$ the transition probability matrix becomes
\begin{align}
\begin{aligned}
\big(\f{A}_{\alpha\beta}^{\ell\ell'jj'}(\f{k},\f{k}',t)\big)_{n,n'} = & \, 4\pi^{2} V_{\f{kk}'}^{2}\sum_{\gamma} c_{\alpha\gamma}^{\ell-\ell'+n}(\f{k},\f{k}')\,\delta\big( \varepsilon-\varepsilon_{\gamma}(\f{k}')-(\ell-\ell')\hbar\Omega \big) \\ &\times \,  \big( c_{\beta\gamma}^{j-j'+n'}(\f{k},\f{k}')\big)^{*}\,\delta\big(\varepsilon-\varepsilon_{\gamma}(\f{k}')-(j-j')\hbar\Omega \big)\ .
\end{aligned}
\end{align}
Since the quasienergies are always defined
to be in the central
Floquet zone, c.f. Eq.~\eqref{FermiProductDelta}, the probability
matrix simplifies to 
\begin{align}
\big(\f{A}_{\alpha\beta}^{\ell\ell'jj'}(\f{k},\f{k}',t)\big)_{n,n'} = 4\pi^{2}V^{2}_{\f{kk}'} \sum_{\gamma}  c_{\alpha\gamma}^{n}(\f{k},\f{k'}) \big(c_{\beta\gamma}^{n'}(\f{k},\f{k'})\big)^{*} \,\delta^{2} \big( \varepsilon-\varepsilon_{\gamma}(\f{k}') \big)\ .
\end{align}
Using Eq.~\eqref{FermiDeltaSquared} and performing the time derivative
of each matrix element yields
\begin{align}
\Gamma_{\alpha\beta}^{nn'}(\f{k},\f{k}') & \equiv \frac{d\big(\f{A}_{\alpha\beta}^{\ell\ell'jj'}(\f{k},\f{k}',t)\big)_{n,n'}}{dt} \\
& = \frac{2\pi}{\hbar}V_{\f{kk}'}^{2} \sum_{\gamma}  c_{\alpha\gamma}^{n}(\f{k},\f{k'}) \big(c_{\beta\gamma}^{n'}(\f{k},\f{k'}) \big)^{*} \,\delta \big( \varepsilon-\varepsilon_{\gamma}(\f{k}') \big)\;.
\end{align}
Finally, one can perform an impurity average and identify
$\langle V^{2}_{\f{kk}'}\rangle_{\mathrm{imp}} =\Vimp$. Summing the
rate over all momenta one gets the inverse
scatting time matrix,  
\begin{align} \label{ScatteringTimett'states}
\left(\frac{1}{\bs{\tau}(\varepsilon,\f{k})} \right)_{\alpha\beta}^{nn'} &\equiv  \frac{1}{V_{\f{k}'}}\sum_{\f{k}'} \langle \Gamma_{\alpha\beta}^{nn'}(\f{k},\f{k}')\rangle_{\mathrm{imp}} \\
&= \frac{2\pi}{\hbar} \Vimp \frac{1}{V_{\f{k}'}}\sum_{\f{k}'} \sum_{\gamma}  c_{\alpha\gamma}^{n}(\f{k},\f{k'}) \big(c_{\beta\gamma}^{n'}(\f{k},\f{k'}) \big)^{*} \,\delta \big( \varepsilon-\varepsilon_{\gamma}(\f{k}') \big) \\
& = i \Big( \bs{T}^{\dagger}(\f{k}) \big( \se^{r}_{1\mathrm{BA}}(\varepsilon,\f{k}) - \se^{a}_{1\mathrm{BA}}(\varepsilon,\f{k})\big)\bs{T}(\f{k}) \Big)_{\alpha\beta}^{nn'}\;.
\end{align}
This expression is equal to the result derived from the Dyson series for the Floquet
Green's function. Remarkably, the central entry of the scattering time
for the $tt'$ Floquet states is equal to the Floquet Fermi's golden rule given in
Eq.~\eqref{ScatteringTimeGoldenRule} and
Refs.~\cite{Kitagawa11,Bilitewski15}.
\section{Definition of the Floquet Zone}
\begin{figure}[h]
	\centering
	\includegraphics[width=0.4\columnwidth]{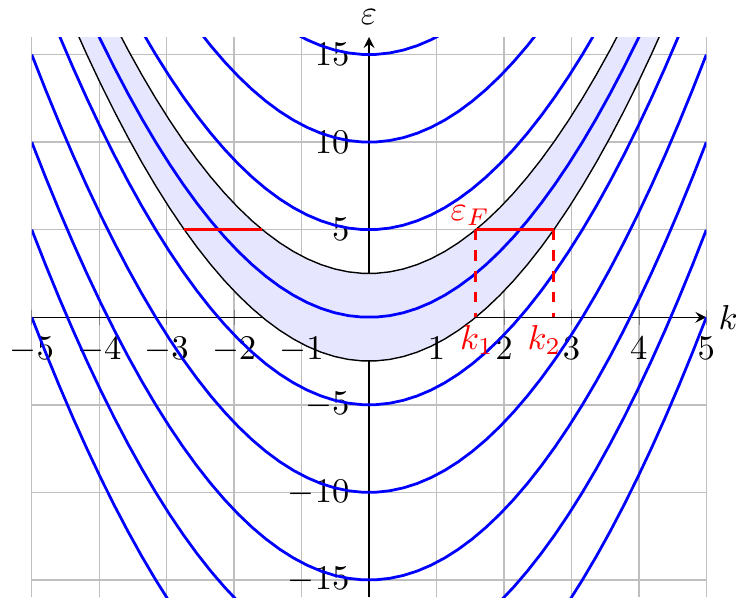}
	\caption{The Floquet zones are chosen to wrap around the
          parabolas. The quasi-Fermi energy is only defined in a
          certain momentum range, i.e., $k\in [k_{1},k_{2})$, in the
          central Floquet zone.}\label{FloquetParabola2}
\end{figure}

\begin{figure}[h]
	\centering
	\includegraphics[width=0.4\columnwidth]{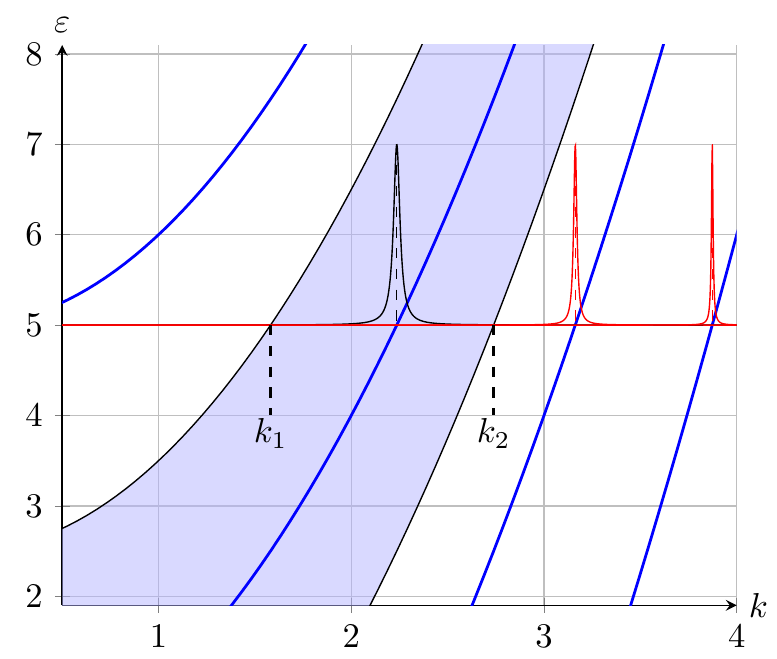}
	\caption{The peaks show the broadening of the Floquet bands
          caused by the scattering time. The blue shaded area is the
          central Floquet zone. If $\Omega\tau_{0}\gg 1$, the leaking
          of the nonzero Floquet modes (red curves) into the central
          Floquet zone is negligibly small.}\label{Tau_no_leaking}
\end{figure}

In the following we would like focus on the a parabolic spectrum and
describe the appropriate choice of the function $\lambda$, which
defines the boundary for the quasienergy $\varepsilon_{\alpha}$,
\begin{align}
  \forall_\alpha: \lambda - \frac{\hbar\Omega}{2} \le \varepsilon_{\alpha} < \lambda + \frac{\hbar\Omega}{2}\;.
\end{align}
Since the spectrum is not bounded, one has to choose the Floquet zone
as indicated in Fig.~\ref{FloquetParabola2}.
This limits the validity of the calculation to the
$\Omega\tau_0 \gg 1$ regime as will be clear in the
following. However, this limitation is only a peculiarity of the
unbounded spectrum: In the derivation of the main text we defined the
quasienergies to fulfill
\begin{align}
  \forall_{\vb{k}, \vb{k}'}\qq{:} |\varepsilon_\alpha(\vb{k}) - \varepsilon_\beta(\vb{k}')| <
\hbar\Omega\;.\label{eepOmega}
\end{align}
As a consequence, in a system with a single band the
condition forces the band width to be smaller than
$\hbar\Omega$. Obviously this cannot be fulfilled by the parabolic
spectrum. In the later case, the momentum range where the quasi-Fermi energy is
defined has to be truncated, as depicted with a red line in
Fig.~\ref{FloquetParabola2}: $k_{1}$ and $k_{2}$, are
functions of the driving frequency $\Omega$. For decreasing $\Omega$
the momenta $k_{1}$ and $k_{2}$ move closer together. If the momentum
range $k\in [k_{1},k_{2})$ is of the order of the broadening of the
Green's function, the truncation leads to an incorrect result for the
conductivity. If $\Omega\tau_{0}\gg 1$, $\tau_{0}$ being the scattering time of the
undriven system, the broadening of the nonzero Floquet modes is small
enough such that the leaking into the central Floquet zone is
negligibly small, compare Fig.~\ref{Tau_no_leaking}.
\begin{figure}[h]
	\centering
	\includegraphics[width=0.4\columnwidth]{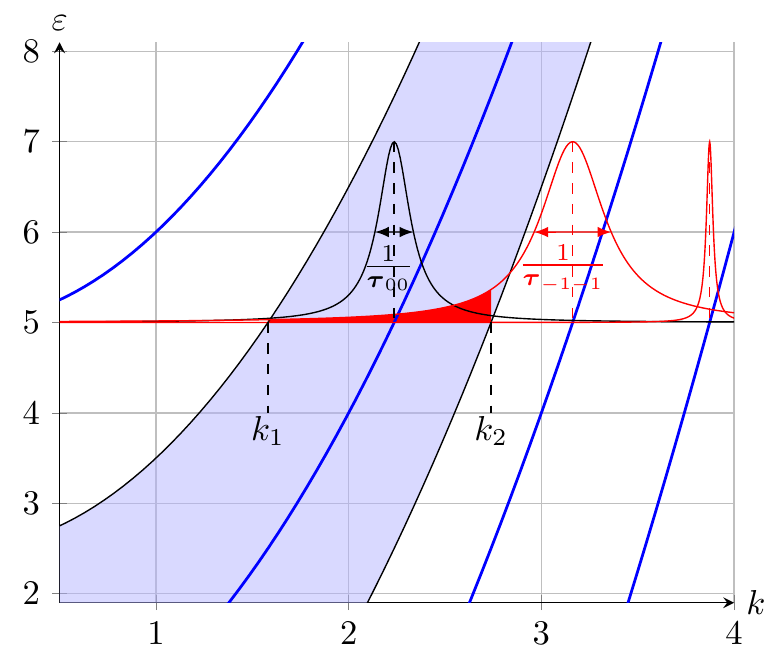}	
 	\caption{The peaks show the broadening of the Floquet bands
          caused by the scattering time. The blue shaded area is the
          central Floquet zone. If $\Omega\tau_{0}\simeq 1$, the
          nonzero Floquet modes are leaking into the central Floquet
          zone. The red shaded area marks the contribution of the
          minus one Floquet band to the
          conductivity.}\label{Tau_leaking}
\end{figure}
If $\Omega\tau_{0}\simeq 1$, the nonzero modes give a significant
contribution to the conductivity as depicted in
Fig.~\ref{Tau_leaking}. In a system which rigorously fulfills
Eq.~\eqref{eepOmega} this constitutes no limitation. 
\section{Example: Driven Square Lattice}
In this section we derive a fully analytic solution of the time-dependent
Schr\"odinger equation for the quadratic lattice with time-periodic
driving and compare it with the results for the parabolic
dispersion. We define the two lattice vectors of the square lattice with
a lattice constant $a$ as
\begin{align}
   \mathbf{a}_{1} = a\begin{pmatrix} 1 \\ 0 \end{pmatrix}  \quad , \quad
   \mathbf{a}_{2} = a\begin{pmatrix} 0 \\ 1 \end{pmatrix}\;.
\end{align}
We choose the vector potential as
\begin{align}
   \mathbfcal{A} = \begin{pmatrix} \mathcal{A}_{x} \sin(\Omega t) \\ \mathcal{A}_{y}\cos(\Omega t) \end{pmatrix}
\end{align}
which allows us to tune the polarization between linear, elliptic and
circular for an appropriate choices of amplitudes $\mathcal{A}_i$. The
time-dependent tight-binding Hamiltonian with hopping parameter $g$
and a limitation to nearest neighbour hopping is this given by
\begin{align}
  \begin{aligned} \label{HtSquare}
    H(t) = -g\Big[ e^{i\f{k}\cdot\f{a}_{1}}\cdot e^{i\frac{e}{\hbar}\mathbfcal{A}\cdot\f{a}_{1}} + e^{i\f{k}\cdot\f{a}_{2}}\cdot e^{i\frac{e}{\hbar}\mathbfcal{A}\cdot\f{a}_{2}} + \text{H.C.} \Big]\ .
\end{aligned}
\end{align}
In the following we will make use of the identities
\begin{align}
  \int dt\, e^{i\gamma \sin(\Omega t)} ={}&  \sum_{n=-\infty}^{\infty} J_{n}(\gamma) \int dt\,
                                            e^{in\Omega t} = \sum_{n\ne 0} \frac{J_{n}(\gamma)}{in\Omega}
                                            e^{in\Omega t} +
                                            J_{0}(\gamma)t\;,\\
  \int dt\, e^{i\gamma \cos(\Omega t)} ={}&  2\sum_{n =1}^{\infty} \frac{i^{n}J_{n}(\gamma)}{n\Omega} \sin(n\Omega t) + J_{0}(\gamma)t\;,
\end{align}
which are based on the Jacobi-Anger expansion.
In order to solve the time-dependent Schr\"odinger equation
\begin{align}
i\hbar\frac{\partial}{\partial t} \psi_{\f{k}}(t) = H(t)\psi_{\f{k}}(t)
\end{align}
we choose the ansatz
\begin{align}
   \psi_{\f{k}}(t) = e^{-\frac{i}{\hbar} F(t)}\qq{with} F(t) = \int dt\, H(t)\ .
\end{align}
Integrating the Hamiltonian \eqref{HtSquare} yields
\begin{align}
\begin{aligned}
   F(t) = &\, -2g\big[J_{0}(\gamma_{x})\cos(k_{x}a) + J_{0}(\gamma_{y})\cos(k_{y}a)\big]t  \\
   & -g e^{ik_{x}a}\sum_{n\ne 0} \frac{J_{n}(\gamma_{x})}{in\Omega} e^{in\Omega t} -g e^{-ik_{x}a}\sum_{n\ne 0} \frac{J_{n}(\gamma_{x})}{-in\Omega} e^{-in\Omega t} \\
   & -2ge^{ik_{y}a} \sum_{n =1}^{\infty} \frac{i^{n}J_{n}(\gamma_{y})}{n\Omega} \sin(n\Omega t) -2g e^{-ik_{y}a} \sum_{n =1}^{\infty} \frac{(-i)^{n}J_{n}(\gamma_{y})}{n\Omega} \sin(n\Omega t)\ .
\end{aligned}
\end{align}
The light parameters are defined by  $\gamma_{i} = e\mathcal{A}_{i}/\hbar$.
The quasienergy is the non-oscillatory part of $F(t)$, thus
\begin{align}
   \epsilon = -2g\big[J_{0}(a\gamma_{x})\cos(k_{x}a) + J_{0}(a\gamma_{y})\cos(k_{y}a)\big] \ .
\end{align}
To make contact with the parabolic spectrum we expand around the $\Gamma$ point, 
\begin{align}
  \epsilon_{T} ={}&  -2g\big[ J_{0}(a\gamma_{x}) + J_{0}(a\gamma_{y}) \big] +
                    \frac{\hbar^2\big(J_{0}(a\gamma_{x})k_{x}^{2}+J_{0}(a\gamma_{y})k_{y}^{2}\big)}{2m}\\
  ={}& -2g\big[ J_{0}(a\gamma_{x}) + J_{0}(a\gamma_{y}) \big] +
                    \frac{\hbar^2k_x^2}{2m_{\mathrm{d}}(\gamma_x)} + \frac{\hbar^2k_y^2}{2m_{\mathrm{d}}(\gamma_y)}\label{epsilonT}
\end{align}
where the effective mass $m = \hbar^2/(2ga^2)$ of the undriven system
got renormalized due to the driving,
$m_{\mathrm{d}}(\gamma_i) = m/J_{0}(a\gamma_i)$.
%
It is worth noticing that if the light parameter $\gamma_i$ is close to a
zero of the Bessel function the mass $m_{\mathrm{d},i}$ diverges and
the expansion around the $\Gamma$ point is not applicable
anymore. Moreover, for rather high intensities this shift of the
parabola in Eq.~\eqref{epsilonT} can significantly change the relative
position to the Fermi energy. To go beyond a small renormalization of
the static effective mass, i.e., if $a\gamma_i$ is close to the zero
of $J_0$ or even larger, $a\gamma_i \gtrsim 2.4$, a realistic multiband
model has to be applied.
\bibliographystyle{apsrev4-1}
\bibliography{paper}